{}
{}

\documentclass[12pt,a4paper]{article}

\newif\ifpublic\publictrue

\usepackage{amsmath,amssymb}
\ifpublic\else\usepackage{showkeys}\fi
\usepackage[bookmarks=true,hyperfigures=true]{hyperref}
\usepackage{graphicx}
\usepackage[nosort]{cite}
\usepackage[bulletsep]{collref}

\def\showkeysrefformat#1{{\normalfont\tiny\ttfamily#1}}
\makeatletter
\def\SK@@ref#1>#2\SK@{%
 {\@inlabelfalse\leavevmode\vbox to\z@{%
 \vss\SK@refcolor\rlap{\vrule\raise .75em%
  \hbox{\showkeysrefformat{#2}}}}}}
\makeatother

\usepackage[a4paper,text={160mm,247mm},centering]{geometry}

\usepackage{color}
\newcommand{\remark}[2][.]{{\color{red}\renewcommand{\bfdefault}{b}\rmfamily\if#1.\else\textbf{#1:} \fi#2}}

\allowdisplaybreaks[3]

\numberwithin{equation}{section}

\usepackage[font=small,labelfont=bf,width=0.85\textwidth]{caption}

\expandafter\def\expandafter\bfseries\expandafter{\bfseries\ifmmode\else\boldmath\fi}
\expandafter\def\expandafter\mdseries\expandafter{\mdseries\ifmmode\else\unboldmath\fi}
\expandafter\def\expandafter\normalfont\expandafter{\normalfont\ifmmode\else\unboldmath\fi}

\RequirePackage{verbatim}

\makeatletter
\newwrite\bibinl@out
\newenvironment{bibtex}[1][\jobname]{%
  \immediate\openout\bibinl@out #1.bib
  \immediate\write\bibinl@out{\@percentchar generated from `\jobname' starting line \the\inputlineno^^J}%
  \def\verbatim@processline{\immediate\write\bibinl@out{\the\verbatim@line}}%
  \@bsphack\let\do\@makeother\dospecials\catcode`\^^M\active\verbatim@start
}%
{\immediate\closeout\bibinl@out\@esphack}
\makeatother

\RequirePackage{verbatim}

\makeatletter
\newwrite\mpi@out
\def\mpi@write#1{\immediate\write\mpi@out{#1}}
\immediate\openout\mpi@out\jobname.mp
\mpi@write{\@percentchar generated from `\jobname'}%
\AtEndDocument{\mpostdone}

\def\mpostdone{
  \immediate\closeout\mpi@out%
  \ifpublic\else%
    \immediate\write18{mpost -tex=latex \jobname.mp}
  \fi%
  \gdef\mpostdone{}
}

\newcommand{\mpi@putlineno}{%
  \mpi@write{\@percentchar---------------------------------------}%
  \mpi@write{\@percentchar l.\the\inputlineno}%
}

\newcommand{\mpi@verbatim}{
  \@bsphack
  \let\do\@makeother\dospecials
  \catcode`\^^M\active
  \def\verbatim@processline{\mpi@write{\the\verbatim@line}}%
  \verbatim@start
}

\newenvironment{mpostcmd}{%
  \mpi@putlineno%
  \mpi@verbatim%
}%
{\mpi@write{}\@esphack}

\newenvironment{mpostfile}[1]{%
  \mpi@putlineno%
  \mpi@write{filenametemplate "#1";}%
  \mpi@write{beginfig(0)}%
  \mpi@verbatim%
}%
{\mpi@write{endfig;}\@esphack}

\newcommand{\includegraphicsex}[2][]{%
  \xdef\mpi@tmp{#2}%
  \IfFileExists{\mpi@tmp}%
    {\includegraphics[#1]{\mpi@tmp}}%
    {\textbf{??}\typeout{file \mpi@tmp{} missing}}%
}

\makeatother

\makeatletter
\newsavebox{\apb@box}\newlength{\apb@width}
\newcommand{\autoparbox}[2][c]{\sbox{\apb@box}{#2}%
 \settowidth{\apb@width}{\usebox{\apb@box}}%
 \parbox[#1]{\apb@width}{\usebox{\apb@box}}}
\newcommand{\includegraphicsbox}[2][]{\autoparbox{\includegraphicsex[#1]{#2}}}
\makeatother

\usepackage{fixmath}

\newcommand{\sfrac}[2]{{\textstyle\frac{#1}{#2}}}
\newcommand{\half}{\sfrac{1}{2}}

\newcommand{\Vectors}{\mathbb{V}}

\newcommand{\hopf}[1]{\mathrm{#1}}

\newcommand{\alg}[1]{\mathfrak{#1}}

\newcommand{\gen}[1]{\mathbb{#1}{}}
\newcommand{\rep}{\rho}
\newcommand{\copro}{\mathrm{\Delta}}

\newcommand{\tmat}{\mathcal{T}}
\newcommand{\rmat}{\mathcal{R}}

\newcommand{\rbody}{\mathcal{H}}
\newcommand{\embed}{\mathcal{E}}
\newcommand{\fuse}{\mathcal{F}}

\newcommand{\coproop}{\copro^{\text{op}}}

\newcommand{\fund}{\mathrm{F}}
\newcommand{\bifund}{\mathrm{B}}
\newcommand{\Ufund}{U}
\newcommand{\ridx}[4]{{}^{#1}{}_{#2}{}^{#3}{}_{#4}}

\newcommand{\embedcomp}{\overline{\embed}}
\newcommand{\fusecomp}{\overline{\fuse}}
\newcommand{\bifundcomp}{{\overline{\bifund}}}
\newcommand{\combinecomp}[1]{\combine{\overline{#1}}}

\newcommand{\Spectral}{\mathrm{M}}
\newcommand{\combine}[1]{{\langle #1\rangle}}

\newcommand{\indup}[1]{_{\mathrm{#1}}}

\RequirePackage{amsmath}
\makeatletter
\newcommand{\brk@ord}{\bBigg@{0}}
\newcommand{\brk@ordl}{\mathopen\brk@ord}
\newcommand{\brk@ordr}{\mathclose\brk@ord}
\newcommand{\brk@ordm}{\mathrel\brk@ord}
\newcommand{\brk@var}{\brk@ord}
\newcommand{\brk@varl}{\left}
\newcommand{\brk@varr}{\right}
\newcommand{\brk@varm}{\mathrel\brk@var}
\newcommand{\brk@altname}[3]{\expandafter\def\csname#2\expandafter\@gobble\string#1\endcsname{#1[#3]}}
\newcommand{\brk@usearg}[3]{%
  \def\brk@star{*}\def\brk@blank{}\def\brk@arg{#1}%
  \ifx\brk@arg\brk@blank\def\brk@arg{brk@ord}\fi%
  \ifx\brk@arg\brk@star\def\brk@arg{brk@var}\fi%
  \csname\brk@arg #2\endcsname#3}

\newcommand{\DeclareMathBrackets}[3]{
  \newcommand{#1}[2][]{\brk@usearg{##1}{l}{#2}##2\brk@usearg{##1}{r}{#3}}
  \brk@altname{#1}{big}{big}\brk@altname{#1}{lr}{*}}
\newcommand{\DeclareMathBiBrackets}[4]{
  \newcommand{#1}[3][]{\brk@usearg{##1}{l}{#2}##2#3##3\brk@usearg{##1}{r}{#4}}
  \brk@altname{#1}{big}{big}\brk@altname{#1}{lr}{*}}
\newcommand{\DeclareMathBiMBracketsStar}[4]{
  \newcommand{#1}[3][]{\brk@usearg{##1}{l}{#2}##2\brk@usearg{##1}{m}{#3}##3\brk@usearg{##1}{r}{#4}}
  \brk@altname{#1}{bi}{big}}
\newcommand{\DeclareMathBiBracketsStar}[4]{
  \newcommand{#1}[3][]{\brk@usearg{##1}{l}{#2}##2\brk@usearg{##1}{}{#3}##3\brk@usearg{##1}{r}{#4}}
  \brk@altname{#1}{big}{big}}

\makeatother

\DeclareMathBrackets{\brk}{(}{)}
\DeclareMathBrackets{\sbrk}{[}{]}
\DeclareMathBrackets{\set}{\{}{\}}
\DeclareMathBrackets{\abs}{|}{|}
\DeclareMathBrackets{\eval}{.}{|}
\DeclareMathBrackets{\bra}{\langle}{|}
\DeclareMathBrackets{\ket}{|}{\rangle}
\DeclareMathBrackets{\state}{|}{\rangle}
\DeclareMathBrackets{\spn}{\langle}{\rangle}
\DeclareMathBiBrackets{\comm}{[}{,}{]}
\DeclareMathBiBrackets{\acomm}{\{}{,}{\}}
\DeclareMathBiBrackets{\gcomm}{[}{,}{\}}

\DeclareMathOperator{\tr}{tr}
\DeclareMathOperator{\diag}{diag}

\newcommand{\trans}{{\mathsf{T}}}

\newcommand{\nln}{\nonumber\\}

\def\[{\begin{equation}}
\def\]{\end{equation}}

\providecommand{\href}[2]{#2}

\makeatletter
\def\mr@ignsp#1 {\ifx\:#1\@empty\else #1\expandafter\mr@ignsp\fi}%
\newcommand{\multiref}[1]{\begingroup
\xdef\mr@no@sparg{\expandafter\mr@ignsp#1 \: }%
\def\mr@comma{}%
\@for\mr@refs:=\mr@no@sparg\do{\mr@comma\def\mr@comma{,}\ref{\mr@refs}}%
\endgroup}
\renewcommand{\eqref}[1]{(\multiref{#1})}
\makeatother

\makeatletter
\newcommand{\namedref}[2]{\hyperref[#2]{#1~\ref*{#2}}}
\newcommand{\secref}{\@ifstar{\namedref{Section}}{\namedref{Sec.}}}
\newcommand{\appref}{\@ifstar{\namedref{Appendix}}{\namedref{App.}}}
\newcommand{\tabref}{\@ifstar{\namedref{Table}}{\namedref{Tab.}}}
\newcommand{\figref}{\@ifstar{\namedref{Figure}}{\namedref{Fig.}}}
\makeatother

\let\oldbib=\thebibliography
\def\thebibliography{\phantomsection\addcontentsline{toc}{section}{\refname}\oldbib}

\let\oldtoc=\tableofcontents
\def\tableofcontents{\phantomsection\addcontentsline{toc}{section}{\contentsname}\oldtoc}

\providecommand{\hypersetup}[1]{}

\hypersetup{plainpages=false}
\hypersetup{pdfpagemode=UseNone}
\hypersetup{bookmarksnumbered=true}
\hypersetup{pdfstartview=FitH}
\hypersetup{colorlinks=false}
\hypersetup{citebordercolor={.5 1 .5}}
\hypersetup{urlbordercolor={.5 1 1}}
\hypersetup{linkbordercolor={1 .7 .7}}

\makeatletter
\let\@keywords\@empty
\let\@subject\@empty
\providecommand{\keywords}[1]{\gdef\@keywords{#1}}
\providecommand{\subject}[1]{\gdef\@subject{#1}}
\def\thetitle{\@title}
\def\theauthor{\@author}
\def\thesubject{\@subject}
\def\thedate{\@date}
\def\thekeywords{\@keywords}
\makeatother
\AtBeginDocument{
\hypersetup{pdftitle={\thetitle}}%
\hypersetup{pdfauthor={\theauthor}}%
\hypersetup{pdfsubject={\thesubject}}%
\hypersetup{pdfkeywords={\thekeywords}}%
}

\begin{mpostcmd}
prologues:=2;

verbatimtex
\documentclass{article}
\usepackage{amsfonts,amssymb,latexsym}
\begin{document}
etex
\end{mpostcmd}

\begin{mpostcmd}
picture copyrightline,copyleftline;
copyrightline := btex \copyright\ \textsf{2015 Niklas Beisert} etex;
copyleftline := btex $\circledast$ etex;
def putcopyspace =
label.bot(btex \vphantom{gA} etex scaled 0.1, lrcorner(currentpicture));
enddef;
def putcopy =
label.ulft(copyrightline scaled 0.1, lrcorner(currentpicture)) withcolor 0.9white;
label.urt(copyleftline scaled 0.1, llcorner(currentpicture)) withcolor 0.9white;
currentpicture:=currentpicture shifted (10.5cm,14cm);
enddef;
\end{mpostcmd}

\begin{mpostcmd}
def pensize(expr s)=withpen pencircle scaled s enddef;
xu:=1cm;
pair pos[];
path paths[];
def midarrow (expr p, t) =
fill arrowhead subpath(0,arctime(arclength(subpath (0,t) of p)+0.5ahlength) of p) of p;
enddef;
\end{mpostcmd}

\begin{mpostfile}{FigFuseR1.mps}
paths[1]:=(-1.5xu,-0.5xu){dir 0}..{dir 0}(+1.5xu,+0.5xu);
paths[2]:=(-1.5xu,+0.5xu){dir 0}..{dir 0}(+1.5xu,-0.5xu);
draw paths[1] pensize(1pt);
draw paths[2] pensize(1.5pt);
pos[1]:=paths[1] intersectionpoint paths[2];
fill fullcircle scaled 0.7xu shifted pos[1] withcolor 0.9white;
draw fullcircle scaled 0.7xu shifted pos[1];
label(btex $\mathcal{R}$ etex, pos[1]);
ahlength:=5pt;
midarrow(reverse(paths[1]),0.2);
midarrow(reverse(paths[1]),0.8);
ahlength:=7pt;
midarrow(reverse(paths[2]),0.2);
midarrow(reverse(paths[2]),0.8);
label.lft(btex $\scriptstyle 3$ etex, point 0 of paths[1]);
label.rt(btex $\scriptstyle 3$ etex, point 1 of paths[1]);
label.lft(btex $\scriptstyle \langle 12\rangle$ etex, point 0 of paths[2]);
label.rt(btex $\scriptstyle \langle 12\rangle$ etex, point 1 of paths[2]);
putcopyspace;putcopy;
\end{mpostfile}

\begin{mpostfile}{FigFuseR2.mps}
a:=0.2;
paths[1]:=(-2.5xu,-0.5xu){dir 0}..{dir 0}(+2.5xu,+0.5xu);
paths[2]:=(-2.5xu,+0.5xu){dir 0}..{dir 0}(+2.5xu,-0.5xu);
pos[1]:=point a of paths[2];
pos[2]:=point (1-a) of paths[2];
paths[3]:=pos[1]{dir -60}..{dir 0}pos[2];
paths[4]:=pos[1]{dir 0}..{dir -60}pos[2];
draw paths[1] pensize(1pt);
draw subpath(0,a) of paths[2] pensize(1.5pt);
draw subpath(1-a,1) of paths[2] pensize(1.5pt);
draw paths[3] pensize(0.5pt);
draw paths[4] pensize(0.5pt);
pos[3]:=paths[1] intersectionpoint paths[3];
pos[4]:=paths[1] intersectionpoint paths[4];
fill fullcircle scaled 0.5xu shifted pos[1] withcolor 0.8white;
draw fullcircle scaled 0.5xu shifted pos[1];
label(btex $\mathcal{F}$ etex, pos[1]);
fill fullcircle scaled 0.5xu shifted pos[2] withcolor 0.8white;
draw fullcircle scaled 0.5xu shifted pos[2];
label(btex $\mathcal{E}$ etex, pos[2]);
fill fullcircle scaled 0.6xu shifted pos[3] withcolor 0.9white;
draw fullcircle scaled 0.6xu shifted pos[3];
label(btex $\mathcal{R}$ etex, pos[3]);
fill fullcircle scaled 0.6xu shifted pos[4] withcolor 0.9white;
draw fullcircle scaled 0.6xu shifted pos[4];
label(btex $\mathcal{R}$ etex, pos[4]);
ahlength:=5pt;
midarrow(reverse(paths[1]),0.2);
midarrow(reverse(paths[1]),0.5);
midarrow(reverse(paths[1]),0.8);
ahlength:=7pt;
midarrow(subpath(a,0) of paths[2],0.7);
midarrow(subpath(1,1-a) of paths[2],0.3);
ahlength:=4pt;
midarrow(reverse(paths[3]),0.3);
midarrow(reverse(paths[3]),0.85);
midarrow(reverse(paths[4]),0.15);
midarrow(reverse(paths[4]),0.7);
label.lft(btex $\scriptstyle 3$ etex, point 0 of paths[1]);
label.rt(btex $\scriptstyle 3$ etex, point 1 of paths[1]);
label.lft(btex $\scriptstyle \langle 12\rangle$ etex, point 0 of paths[2]);
label.rt(btex $\scriptstyle \langle 12\rangle$ etex, point 1 of paths[2]);
label.bot(btex $\scriptstyle 1$ etex, point 0.7 of paths[3]);
label.top(btex $\scriptstyle 2$ etex, point 0.3 of paths[4]);
label.llft(btex $\scriptstyle 1$ etex, point 0.15 of paths[3]);
label.urt(btex $\scriptstyle 2$ etex, point 0.85 of paths[4]);
putcopyspace;putcopy;
\end{mpostfile}

\begin{mpostfile}{FigFuseTitle.mps}
a:=0.2;
paths[1]:=(-2.5xu,-0.5xu){dir 0}..{dir 0}(+2.5xu,+0.5xu);
paths[2]:=(-2.5xu,+0.5xu){dir 0}..{dir 0}(+2.5xu,-0.5xu);
pos[1]:=point a of paths[2];
pos[2]:=point (1-a) of paths[2];
paths[3]:=pos[1]{dir -60}..{dir 0}pos[2];
paths[4]:=pos[1]{dir 0}..{dir -60}pos[2];
draw paths[1] pensize(1pt);
draw subpath(0,a) of paths[2] pensize(1.5pt);
draw subpath(1-a,1) of paths[2] pensize(1.5pt);
draw paths[3] pensize(0.5pt);
draw paths[4] pensize(0.5pt);
pos[3]:=paths[1] intersectionpoint paths[3];
pos[4]:=paths[1] intersectionpoint paths[4];
fill fullcircle scaled 0.5xu shifted pos[1] withcolor 0.8white;
draw fullcircle scaled 0.5xu shifted pos[1];
fill fullcircle scaled 0.5xu shifted pos[2] withcolor 0.8white;
draw fullcircle scaled 0.5xu shifted pos[2];
fill fullcircle scaled 0.6xu shifted pos[3] withcolor 0.9white;
draw fullcircle scaled 0.6xu shifted pos[3];
fill fullcircle scaled 0.6xu shifted pos[4] withcolor 0.9white;
draw fullcircle scaled 0.6xu shifted pos[4];
putcopyspace;putcopy;
\end{mpostfile}

\begin{mpostcmd}
verbatimtex
\end{document}
etex

end;
\end{mpostcmd}

\mpostdone

\title{Fusion for the one-dimensional Hubbard model}
\author{Niklas Beisert, Marius de Leeuw and Panchali Nag}

\begin{document}

\pdfbookmark[1]{Title Page}{title}
\thispagestyle{empty}


\vspace*{2cm}
\begin{center}%
\begingroup\Large\bfseries\thetitle\par\endgroup
\vspace{1cm}

\begingroup\scshape
Niklas Beisert${}^{1}$, Marius de Leeuw${}^{1,2}$ and Panchali Nag${}^{1}$
\par\endgroup
\vspace{5mm}%

\begingroup\itshape
${}^{1}$ Institut f\"ur Theoretische Physik,\\
Eidgen\"ossische Technische Hochschule Z\"urich\\
Wolfgang-Pauli-Strasse 27, 8093 Z\"urich, Switzerland
\vspace{3mm}

${}^{2}$ Niels Bohr Institute,\\
Copenhagen University\\
Blegdamsvej 17, 2100 Copenhagen \O, Denmark

\par\endgroup
\vspace{5mm}

\begingroup\ttfamily
nbeisert@itp.phys.ethz.ch,
deleeuwm@nbi.ku.dk,
nagp@student.ethz.ch\par
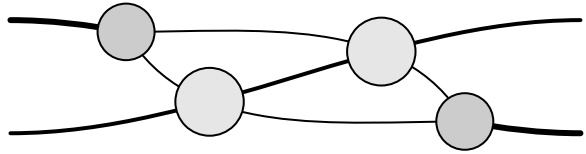
\endgroup

\vfill

\includegraphicsbox[scale=1.5]{FigFuseTitle.mps}

\vfill

\pdfbookmark[1]{Abstract}{abs}
\textbf{Abstract}\vspace{5mm}

\begin{minipage}{12.7cm}
We discuss a formulation of the fusion procedure for integrable models
which is suitable for application to non-standard R-matrices.
It allows for construction of bound state R-matrices
for AdS/CFT worldsheet scattering or equivalently for
the one-dimensional Hubbard model.
We also discuss some peculiar cases that arise in these models.
\end{minipage}

\vspace*{4cm}

\end{center}

\newpage



\section{Introduction}
\label{sec:intro}

Integrable systems constitute a special class of physical models that are exactly solvable \cite{BaxterBook}.
The key ingredient that allows for the explicit construction of
exact solutions is the so-called R-matrix.
For most known integrable models, the corresponding R-matrices
are determined by the underlying symmetry algebra of the system.
This is usually an infinite-dimensional Hopf algebra
of Yangian or quantum affine type.

Computing the R-matrix in various representations of such an algebra
then describes different types of particles.
For example, the Heisenberg XXX model has the Yangian of $\alg{su}(2)$ as its symmetry algebra.
The R-matrix in the fundamental representation
simply describes a chain of spin-$\half$ particles.
Similarly, particles of higher spin can be considered
by taking higher-dimensional representations of $\alg{su}(2)$.

On the other hand, from basic representation theory
it is well known that for example the tensor product of two spin-$\half$ particles
splits into a spin-1 and a spin-0 representation.
In other words, one should be able to relate R-matrices
in higher dimensional representations to the fundamental R-matrix.
For example, the R-matrix of spin-1 particles should allow
for some decomposition into the R-matrix of spin-$\half$ particles.
Similarly, from the fundamental representation
it should be possible to construct new R-matrices corresponding to other representations.
This construction goes under the name of fusion
\cite{Karowski:1978ps, Kulish:1981gi, Jimbo:1985zk, Jimbo:1985vd, Kirillov:1987zz, Mezincescu:1991ke}.
For most integrable systems this procedure is well-understood,
however, the established formulas do not directly apply
for some of the more exotic integrable models.

Recently, there was renewed interest in the field of integrable systems
due to their appearance in string and gauge theory via the AdS/CFT correspondence (
see \cite{Beisert:2010jr} and references therein).
In particular, this integrable structure gave rise to an unusual R-matrix
that displays Yangian symmetry corresponding
to the centrally extended $\alg{su}(2|2)$ Lie superalgebra \cite{Beisert:2005tm, Arutyunov:2006ak, Beisert:2007ds}.
Remarkably, this R-matrix turned out to be directly related to Shastry's R-matrix \cite{ShastryMatrix}
describing the one-dimensional Hubbard model \cite{Beisert:2006qh, Martins:2007hb}.

It soon became clear that for a full description of the string model,
the R-matrices in higher dimensional representations corresponding
to bound states were needed \cite{Dorey:2006dq, Chen:2006gea, Arutyunov:2007tc}.
These bound state R-matrices could be computed directly by invoking
the Yang--Baxter equation \cite{Arutyunov:2008zt} or by using Yangian symmetry \cite{deLeeuw:2008dp,Arutyunov:2009mi}.
However, how to obtain these matrices directly
from the fundamental R-matrix was unknown since the usual fusion procedure breaks down.
In this note we will introduce a slight generalization
of the fusion procedure for integrable models which allows us
to obtain bound state R-matrices for the AdS/CFT S-matrix and Shastry's R-matrix.

This paper is organized as follows.
In \secref*{sec:general} we introduce our fusion procedure
and study it at the level of R-matrices.
Then in \secref*{sec:advanced} we study more advanced aspects of it.
Afterwards we turn to applications:
In \secref*{sec:XXX} we discuss the well-known example
of the XXX spin chain before
moving on to the novel case of AdS/CFT worldsheet scattering
and the one-dimensional Hubbard model
in \secref*{sec:Hubb}.

\section{Fusion}
\label{sec:general}

Consider an integrable system whose fundamental degrees of freedom
are described by an $n$-dimensional vector space $\Vectors^{\fund}$.
These might represent the spin degrees of freedom of an integrable
spin chain or the particle flavors of an integrable scattering problem
in $1+1$ dimensions.
Their interactions are described by an $(n^2\times n^2)$-dimensional R-matrix
\begin{align}
\rmat(u_1,u_2):
\Vectors^{\fund} \otimes \Vectors^{\fund} \rightarrow
\Vectors^{\fund} \otimes \Vectors^{\fund},
\end{align}
where the parameters $u_1,u_2\in \Spectral^\fund$ describe the
inhomogeneities of the spin sites or the
particle rapidities.
The space of parameters $\Spectral^\fund$ is
typically a one-dimensional complex manifold,%
\footnote{Additional continuous parameters or even discrete parameters
are conceivable as well.}
such as the complex plane or the Riemann sphere with certain punctures.

For an integrable system, the R-matrix satisfies the Yang--Baxter equation and
the involution property
(the latter relation is understood up to an overall factor)
\begin{align}\label{eq:YBE}
&\rmat_{12}\rmat_{13}\rmat_{23} =
\rmat_{23}\rmat_{13}\rmat_{12} ,
&& \rmat_{12}\rmat_{21}\sim 1_{12}.
\end{align}
Here and in the following, the indices denote the spaces
in a tensor product as well as the associated parameters $u_k$.

\subsection{Singularities}

Suppose that there are pairs of points
\[
u_\combine{12}=\bigbrk{u_1,u_2}\in \Spectral^\bifund\subset \Spectral^\fund\times\Spectral^\fund,
\]
where the R-matrix becomes non-invertible.%
\footnote{The R-matrix can have singular points
where some of its eigenvalues diverge.
As we are not interested in overall factors of the R-matrix
(which may well depend on $u_1$ and $u_2$)
we should rescale the R-matrix at these points to remove the singularity.
In other words, what counts is
that the leading contribution in a Laurent expansion is non-invertible.}
In other words, the rank of the R-matrix drops below its maximum%
\[
\operatorname{rank} \rmat(u_1,u_2) = m<n^2.
\]
Commonly, these points form
a one-dimensional sub-manifold%
\footnote{Discrete points with this property
can also be considered along the same lines.
We will, however, mostly be interested
in the case of continuous values for $u_\combine{12}$.}
$\Spectral^\bifund$ of $\Spectral^\fund\times\Spectral^\fund$.
The point $u_\combine{12}\in\Spectral^\bifund$ can thus be treated as
a continuous parameter, and it is on the same footing as
the parameters $u_k\in\Spectral^\fund$.
In scattering theory, such singular behavior signals the existence
of composite particles which are naturally part of the physical scattering problem.%
\footnote{In fact, they are only part of the physical problem if they are also \emph{bound}.
The distinction between bound and unbound composite particles
makes no difference here, and we shall refer to them collectively as composite states.}
For these we will define new R-matrices to be interpreted as
scattering matrices involving composite particles.
We can thus extend the integrable system by adding these composite particles
as additional degrees of freedom.

More precisely, we are led to the introduction
of an $m$-dimensional vector space $\Vectors^{\bifund}$
by `fusing' two spaces $\Vectors^{\fund}$
such that the extended integrable system is defined on
$\Vectors^{\text{ext}} := \Vectors^{\fund} \oplus \Vectors^{\bifund}$.
The corresponding R-matrix can be schematically written in block form
\begin{align}
&\rmat^{\text{ext}} :
\Vectors^{\text{ext}}\otimes \Vectors^{\text{ext}}
\rightarrow \Vectors^{\text{ext}}\otimes \Vectors^{\text{ext}},
&& \rmat^{\text{ext}} =
\diag (\rmat^{\fund\fund}, \rmat^{\fund\bifund}, \rmat^{\bifund\fund}, \rmat^{\bifund\bifund}),
\end{align}
where the various blocks are maps of the types ($A_i = \fund,\bifund$)
\[
\rmat^{A_1 A_2}(u_1,u_2) : \Vectors^{A_1}\otimes \Vectors^{A_2}
  \rightarrow \Vectors^{A_1}\otimes \Vectors^{A_2} ,
\qquad
(u_1,u_2)\in \Spectral^{A_1}\times \Spectral^{A_2}.
\]
In particular, $\rmat^{\fund\fund} \equiv \rmat$.
The fact that $\rmat^{\text{ext}}$
satisfies the Yang--Baxter equation is then equivalent to
the statement
\[
\label{eq:generalYBE}
\rmat_{12}^{A_1 A_2}\rmat_{13}^{A_1 A_3}\rmat_{23}^{A_2 A_3} =
\rmat_{23}^{A_2 A_3}\rmat_{13}^{A_1 A_3}\rmat_{12}^{A_1 A_2}.
\]
Moreover, the R-matrices $\rmat^{A_1 A_2}$
have the involution property
\[
\label{eq:generalInvolution}
\rmat_{12}^{A_1 A_2}\rmat_{21}^{A_2 A_1}
\sim 1_{12}^{A_1 A_2}.
\]
Of course the fusion procedure can be recursively applied,
leading to larger and larger integrable systems.

\subsection{Procedure}

In the following, we shall construct the fused R-matrices
and afterwards check their properties.

\paragraph{Embedding and fusion matrices.}

We consider the R-matrix
$\rmat(u_\combine{12}):=\rmat(u_1,u_2)$ at a point $u_\combine{12}$
where the rank drops to $m<n^2$.
At this point we can decompose it as a product of three
matrices
\begin{align}
&\embed(u_\combine{12}): \Vectors^{\bifund} \rightarrow \Vectors^{\fund} \otimes \Vectors^{\fund},
\nln
&\rbody(u_\combine{12}): \Vectors^{\bifund} \rightarrow \Vectors^{\bifund},
\nln
&\fuse(u_\combine{12}): \Vectors^{\fund}\otimes\Vectors^{\fund} \rightarrow \Vectors^{\bifund},
\end{align}
with the properties
\begin{align}\label{eq:PpropertiesA}
&\rmat(u_\combine{12})  =  \embed(u_\combine{12})\, \rbody(u_\combine{12})\, \fuse(u_\combine{12}),
&& \fuse(u_\combine{12})\, \embed(u_\combine{12}) = 1^{\bifund},
&& \rbody\text{ invertible}.
\end{align}
Here, $\fuse$ is a (surjective) $m\times n^2$ matrix which fuses the tensor
product $\Vectors^\fund\otimes\Vectors^\fund$ to the space $\Vectors^\bifund$,
and $\embed$ is an (injective) $n^2\times m$ matrix which embeds $\Vectors^\bifund$
into $\Vectors^\fund\otimes\Vectors^\fund$.
For convenience, we assume these two matrices to be pseudo-inverses as in \eqref{eq:PpropertiesA}.
Finally, $\rbody$ is a (bijective) $m\times m$ matrix and it represents
the action of $\rmat$ on the space $\Vectors^\bifund$.
In the following we shall describe two ways of obtaining the decomposition.

First of all, let $\Vectors^\bifund$ be the image of $\rmat$.
Thus $\rmat$ can be understood as a map
$\Vectors^\fund\otimes\Vectors^\fund\to\Vectors^\bifund$.
Furthermore, define $\embed$ as the trivial embedding map
$\Vectors^\bifund\to\Vectors^\fund\otimes\Vectors^\fund$.
The above decomposition can be obtained as
$\fuse=(\rmat\embed)^{-1}\rmat$ and
$\rbody=\rmat\embed$.
Note that this construction requires $\rbody=\rmat\embed$ to be invertible.
This combination is not invertible precisely if the
image of $\rmat$ contains a vector that is in the kernel of $\rmat$.
In other words, the map $\rmat$ contains a non-trivial nilpotent part.
This case is more difficult to handle, and we shall exclude it for the time being.
Later on in \secref{sec:singlet}, we shall discuss an explicit example.

An alternative construction of the decomposition uses eigenvectors
where we assume that the Jordan decomposition is trivial
(i.e.\ the nilpotent case is excluded).
The matrix $\rmat$ possesses $m$ non-zero eigenvalues $\lambda_1,\ldots,\lambda_m$.
Let $e_1,\ldots,e_m$ denote the associated right eigenvectors of $\rmat$
and $e^1,\ldots,e^m$ the left eigenvectors
\begin{align}
&
\rmat e_k = \lambda_k e_k,
&&
e^k \rmat = \lambda_k e^k.
\end{align}
All of these quantities are functions of $u_\combine{12}$.
We normalize the vectors such that they form two dual bases
for the space $\Vectors^\bifund$
\begin{align}
&
e^k e_l = \delta^k_l,
&&
\rmat = \sum\nolimits_k \lambda_k e_k e^k.
\end{align}
We then define $\embed$, $\rbody$ and $\fuse$ as
the matrices of eigenvectors and eigenvalues
\begin{align}
\label{eq:EHFeigen}
\embed &=
\begin{pmatrix}
e_1 & e_2 & \ldots & e_m
\end{pmatrix},
&
\rbody &=
\diag(\lambda_1,\ldots,\lambda_m),
&
\fuse &=
\begin{pmatrix}
e^1 & e^2 & \ldots & e^m
\end{pmatrix}{}^\trans.
\end{align}
By their definition $\embed$ and $\fuse$ satisfy the
relations \eqref{eq:PpropertiesA}.

The relations \eqref{eq:PpropertiesA} imply the following useful identities
\begin{align}\label{eq:Pproperties}
 \embed(u_\combine{12})\, \rbody(u_\combine{12}) &= \rmat(u_\combine{12})\, \embed(u_\combine{12}),
\nln
  \rbody(u_\combine{12})\, \fuse(u_\combine{12})&=  \fuse(u_\combine{12})\, \rmat(u_\combine{12}),
\nln
 \rmat(u_\combine{12})\, \embed(u_\combine{12})\, \fuse(u_\combine{12})   &= \rmat(u_\combine{12}) .
\end{align}
These are the crucial relations that enable us to carry out the fusion procedure.

\paragraph{Fused R-matrices.}

\begin{figure}
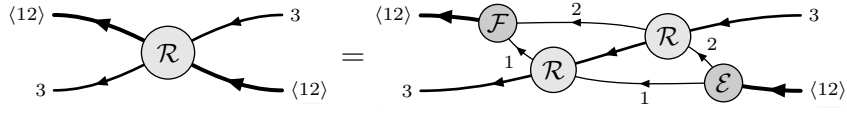
\centering
$\includegraphicsbox{FigFuseR1.mps}
=
\includegraphicsbox{FigFuseR2.mps}$
\caption{Diagrammatical representation of fusion.}
\label{fig:fusion}
\end{figure}

We introduce R-matrices by using $\embed,\fuse$
to `fuse' together two spaces
$\Vectors^{\fund}_{1}\otimes \Vectors^{\fund}_2$
into $\Vectors^{\bifund}_{\combine{12}}$
with
$u_\combine{12}=(u_1,u_2)$.
In particular, we are led to (cf.\ \figref{fig:fusion})
\begin{align}\label{eq:FusedRBF}
\rmat^{\bifund\fund}_{\combine{12}3}(u_{\combine{12}},u_3)  &:=
\fuse_{\combine{12}}(u_{\combine{12}}) \,
\rmat_{13}(u_1,u_3)\,
\rmat_{23}(u_2,u_3)\,
\embed_{\combine{12}}(u_{\combine{12}}),
\\
\label{eq:FusedRFB}
\rmat^{\fund\bifund}_{1\combine{23}}(u_1,u_{\combine{23}})  &:=
\fuse_{\combine{23}}(u_{\combine{23}}) \,
\rmat_{13}(u_1,u_3)\,
\rmat_{12}(u_1,u_2)\,
\embed_{\combine{23}}(u_{\combine{23}}).
\end{align}
Any state from $\Vectors^{\bifund}_{\combine{12}}\otimes \Vectors^{\fund}_3$
is mapped to $\Vectors^{\fund}_1 \otimes \Vectors^{\fund}_{2} \otimes \Vectors^{\fund}_3$
by $\embed$,
acted upon with $\rmat$ and then mapped back by $\fuse$.

Notice the different ordering of the R-matrices
which is needed when spaces two and three are fused rather than spaces one and two.
The R-matrix $\rmat^{\bifund\bifund}$
can then be defined by applying the fusion procedure twice
\begin{align}
\label{eq:FusedRBB}
\rmat^{\bifund\bifund}_{\combine{12}\combine{34}}(u_\combine{12},u_\combine{34})  &:=
\fuse_\combine{34}(u_\combine{34})\,
\rmat^{\bifund\fund}_{\combine{12}4}(u_\combine{12},u_4) \,
\rmat^{\bifund\fund}_{\combine{12} 3}(u_\combine{12},u_3) \,
\embed_\combine{34}(u_\combine{34}).
\end{align}
In particular, from \eqref{eq:Pproperties}
it is readily seen that \eqref{eq:FusedRBB} can be cast into the symmetric form
\begin{align}
\rmat_{\combine{12}\combine{34}}
&:=
\fuse_{\combine{12}}\fuse_{\combine{34}}
\rmat_{14} \rmat_{24} \rmat_{13} \rmat_{23}
\embed_{\combine{12}}\embed_{\combine{34}},
\end{align}
which demonstrates that it is independent
of the order of fusing the underlying spaces.
Here and in the following,
we drop the parameters $u$
and the labels $\fund,\bifund$ in favor of a more concise presentation.
They can be recovered from the labels of the associated spaces.

\subsection{Relations}

In order to show that these R-matrices indeed describe
an integrable system we have to show that they are invertible
and that they satisfy the Yang--Baxter equation.

\paragraph{Involution property.}

We have to show the involution property \eqref{eq:generalInvolution}
which reads more explicitly
\begin{align}
\rmat^{\bifund\fund}_{\combine{12}3}(u_\combine{12},u_3)
\rmat^{\fund\bifund}_{3\combine{12}}(u_3,u_\combine{12}) &\sim
1,
\\
\rmat^{\bifund\bifund}_{\combine{12}\combine{34}}(u_\combine{12},u_\combine{34})
\rmat^{\bifund\bifund}_{\combine{34}\combine{12}}(u_\combine{34},u_\combine{12})
&\sim
1 .
\end{align}
Let us prove the first instance.
For conciseness we will omit the arguments.
Furthermore, we shall put brackets around the terms to be transformed
in the next step
\begin{align}
\rbody_\combine{12}
\rmat_{\combine{12}3}
\rmat_{3\combine{12}}
&=
\bigsbrk{\rbody_\combine{12}\fuse_\combine{12}}\rmat_{13}\rmat_{2 3}\embed_\combine{12}
\fuse_\combine{12}\rmat_{3 2}\rmat_{31}\embed_\combine{12}
\nln
&=
\fuse_\combine{1 2}\bigsbrk{\rmat_{1 2}\rmat_{13}\rmat_{2 3}}
\embed_\combine{1 2}\fuse_\combine{1 2}
\rmat_{3 2}\rmat_{31}\embed_\combine{1 2}
\nln
&=
\fuse_\combine{1 2}
\rmat_{2 3}\rmat_{13}\bigsbrk{\rmat_{1 2}\embed_\combine{1 2}\fuse_\combine{1 2}}
\rmat_{3 2}\rmat_{31}\embed_\combine{1 2}
\nln
&=\fuse_\combine{1 2}\bigsbrk{\rmat_{ 2 3}\rmat_{13}\rmat_{1 2}}
\rmat_{3 2}\rmat_{31}\embed_\combine{1 2}
\nln
&=\bigsbrk{\fuse_\combine{1 2}\rmat_{1 2}}
\bigsbrk{\rmat_{13}\bigsbrk{\rmat_{ 2 3}\rmat_{3 2}}\rmat_{31}}\embed_\combine{1 2}
\nln
&\sim
\rbody_\combine{12}
\bigsbrk{\fuse_\combine{1 2}\embed_\combine{1 2}}
\nln
&=\rbody_\combine{12}.
\end{align}
Invertibility of $\rbody$ then gives the desired result.
In general, the strategy is to remove intermediate factors
of $\embed$ and $\fuse$ by use of the Yang--Baxter equation \eqref{eq:YBE}
and the properties \eqref{eq:Pproperties}.

\paragraph{Yang--Baxter equation.}

Subsequently, it also follows directly from \eqref{eq:Pproperties}
that the above introduced R-matrices \eqref{eq:FusedRBF}--\eqref{eq:FusedRBB}
satisfy the various versions of the Yang--Baxter equation outlined in \eqref{eq:generalYBE}.

For example, since $\rmat$
satisfies the Yang--Baxter equation \eqref{eq:YBE} itself,
\eqref{eq:Pproperties} yields (we again suppress the explicit arguments)
\begin{align}
\rbody_\combine{12}
\rmat_{\combine{12}3}
\rmat_{\combine{12}4}
\rmat_{34}
& =
\bigsbrk{ \rbody_\combine{12} \fuse_\combine{1 2}}
 \rmat_{13} \rmat_{2 3}\embed_\combine{1 2}\fuse_\combine{1 2}
\rmat_{14} \rmat_{2 4}\bigsbrk{\embed_\combine{12} \rmat_{34}}
\nln
& =
\fuse_\combine{1 2}
\bigsbrk{\rmat_{1 2} \rmat_{13} \rmat_{2 3}\embed_\combine{1 2}}\fuse_\combine{1 2}
\rmat_{14} \rmat_{2 4} \rmat_{34}\embed_\combine{12}
\nln
& =
\fuse_\combine{1 2}
\rmat_{1 2} \bigsbrk{\rmat_{13} \rmat_{2 3} \rmat_{14}  \rmat_{2 4} \rmat_{34}}
\embed_\combine{12}
\nln
& =
\bigsbrk{\fuse_\combine{1 2} \rmat_{1 2} \rmat_{34}} \rmat_{14} \rmat_{2 4} \rmat_{13} \rmat_{2 3}
\embed_\combine{12}
\nln
& =
\rmat_{34}
\fuse_\combine{1 2} \bigsbrk{\rmat_{12} \rmat_{14} \rmat_{24}} \rmat_{13} \rmat_{2 3}
\embed_\combine{12}
\nln
& =
\rmat_{34}
\bigsbrk{\fuse_\combine{1 2} \rmat_{12}} \rmat_{14} \rmat_{24} \embed_\combine{12}
\fuse_\combine{12}\rmat_{13} \rmat_{2 3}\embed_\combine{12}
\nln
& =
\bigsbrk{\rmat_{34} \rbody_\combine{12}}
\bigsbrk{\fuse_\combine{1 2} \rmat_{14} \rmat_{24} \embed_\combine{12}}
\bigsbrk{\fuse_\combine{12}\rmat_{13} \rmat_{2 3}\embed_\combine{12}}
\nln
& =
\rbody_\combine{12} \rmat_{34}
\rmat_{\combine{12}4}
\rmat_{\combine{12}3}.
\end{align}
This proves that the R-matrices \eqref{eq:FusedRBF}--\eqref{eq:FusedRBB}
indeed describe the scattering of a new (composite) type of particle in this model.

\section{Further properties}
\label{sec:advanced}

We have established the basic features of fused R-matrices.
In the following we will discuss further properties.

\subsection{Algebra}

Suppose there is a Hopf algebra $\hopf{H}$ describing the symmetries of our integrable system.
In particular, the R-matrix, by definition,
intertwines the coproduct and opposite coproduct in the representation $\rep^{\fund}(u):\hopf{H}\to\Vectors^{\fund}$
under which our fundamental degrees of freedom transform, i.e.\
for any generator $\gen{J}\in \hopf{H}$
\begin{align}\label{eq:symmS}
&\bigbrk{\rep^{\fund}_1(u_1) \otimes \rep^{\fund}_2(u_2)} \bigsbrk{\coproop(\gen{J})} \, \rmat_{12}(u_1,u_2)
= \rmat_{12}(u_1,u_2)\, \bigbrk{\rep^{\fund}_1(u_1) \otimes \rep^{\fund}_2(u_2)}  \bigsbrk{\copro(\gen{J}) }.
\end{align}
We define a new representation
$\rep^{\bifund}(u_\combine{12}):\hopf{H}\to\Vectors^{\bifund}$
for the composite degrees of freedom
\begin{align}\label{eq:DefFusedRep}
&\rep^{\bifund}_\combine{12}(u_\combine{12})\bigsbrk{\gen{J}}
:= \fuse_\combine{12}(u_\combine{12}) \, \bigbrk{\rep^{\fund}_1(u_1) \otimes \rep^{\fund}_2(u_2)}
\bigsbrk{\copro(\gen{J})}\, \embed_\combine{12}(u_\combine{12}).
\end{align}
We will refer to this as the fused or composite representation.
This representation clearly is $m$-dimensional.

Let us show that \eqref{eq:DefFusedRep}
indeed defines a representation by proving that it respects the multiplicative structure.
We have from  \eqref{eq:Pproperties}
and the cocommutativity \eqref{eq:symmS} of the coproduct, that
for any $\gen{J},\gen{J}'\in \hopf{H}$
\begin{align}
\rbody_\combine{12} \rep_\combine{12}[\gen{J}]\rep_\combine{12}[\gen{J}']
&=
\fuse_\combine{12} \rmat_{12} \, \bigbrk{\rep_{1} \otimes \rep_{2}} \bigsbrk{\copro(\gen{J})}\, \embed_\combine{12}
\fuse_\combine{12} \, \bigbrk{\rep_1 \otimes \rep_2} \bigsbrk{\copro(\gen{J}')}\, \embed_\combine{12}
\nln
&=
\fuse_\combine{12} \, \bigbrk{\rep_{1} \otimes \rep_{2}} \bigsbrk{\coproop(\gen{J})}\,
\rmat_{12} \embed_\combine{12} \fuse_\combine{12} \,
\bigbrk{\rep_1 \otimes \rep_2} \bigsbrk{\copro(\gen{J}')}\, \embed_\combine{12}
\nln
&=
\fuse_\combine{12} \, \bigbrk{\rep_{1} \otimes \rep_{2}} \bigsbrk{\coproop(\gen{J})}\,
\rmat_{12} \,
\bigbrk{\rep_1 \otimes \rep_2} \bigsbrk{\copro(\gen{J}')}\, \embed_\combine{12}
\nln
&=
\fuse_\combine{12} \rmat_{12} \, \bigbrk{\rep_{1} \otimes \rep_{2}} \bigsbrk{\copro(\gen{J})} \,
\bigbrk{\rep_1 \otimes \rep_2} \bigsbrk{\copro(\gen{J}')}\, \embed_\combine{12}
\nln
&=
\rbody_\combine{12} \fuse_\combine{12} \, \bigbrk{\rep_1 \otimes \rep_2}
\bigsbrk{\copro(\gen{J}\,\gen{J}')}\, \embed_\combine{12}
\nln
&=\rbody_\combine{12} \rep_\combine{12} [\gen{J}\,\gen{J}'].
\end{align}
Furthermore, the R-matrices \eqref{eq:FusedRBF}--\eqref{eq:FusedRBB}
naturally intertwine the coproduct in the new representation.
Explicitly,
\begin{align}
\bigbrk{\rep^{A_1}_1 \otimes \rep^{A_2}_2} \bigsbrk{\coproop (\gen{J})} \, \rmat^{A_1 A_2}_{12}
&= \rmat^{A_1A_2}_{12}\,\bigbrk{\rep^{A_1}_1 \otimes \rep^{A_2}_2} \bigsbrk{\copro(\gen{J})}.
\end{align}
For instance, let us prove the intertwining relation for the case $A_1=\bifund,$ $A_2=\fund$.
This is most conveniently done in the Sweedler notation
$\copro(\gen{J}) = \sum \gen{J}_{(1)} \otimes \gen{J}_{(2)}$.
Co-associativity of the Hopf algebra is then written as
\begin{align}
\sum \gen{J}_{(1)(1)} \otimes \gen{J}_{(1)(2)} \otimes \gen{J}_{(2)} =
\sum \gen{J}_{(1)} \otimes \gen{J}_{(2)(1)} \otimes \gen{J}_{(2)(2)}
\end{align}
and the intertwining property \eqref{eq:symmS} of the R-matrix is formulated as
\begin{align}\label{eq:coasso}
\sum \rmat_{12}\, \rep_1[\gen{J}_{(1)}]\,  \rep_2[\gen{J}_{(2)}]
=
\sum \rep_1[\gen{J}_{(2)}]\,  \rep_2[\gen{J} _{(1)}]\, \rmat_{12} .
\end{align}
In this language, we have
(for conciseness we suppress the arguments $u,v$ of the R-matrices and the sums)
\begin{align}
\rmat_{\combine{12}3}\bigbrk{\rep_\combine{12} \otimes \rep_3} \bigsbrk{\copro (\gen{J})}
&=
\rmat_{\combine{12}3} \fuse_\combine{12} \,
\rep_1[\gen{J}_{(1)}] \,
\rep_2[\gen{J}_{(2)(1)}] \,
\rep_3[\gen{J}_{(2)(2)}] \embed_\combine{12} \,
\nln
&= \fuse_\combine{12} \rmat_{13}\rmat_{23}\,
\rep_1[\gen{J}_{(1)}] \,
\rep_2[\gen{J}_{(2)(1)}] \,
\rep_3[\gen{J}_{(2)(2)}]
\embed_\combine{12}
\nln
&=\fuse_\combine{12}\,
\rep_1[\gen{J}_{(2)(1)}]\,
\rep_2[\gen{J}_{(2)(2)}]\,
\rep_3[\gen{J}_{(1)}]
\embed_\combine{12}\fuse_\combine{12}\rmat_{13}\rmat_{23} \embed_\combine{12}
\nln
&= \bigbrk{\rep_\combine{12} \otimes \rep_3} \bigsbrk{\coproop (\gen{J})} \, \rmat_{\combine{12}3},
\end{align}
where we used \eqref{eq:Pproperties} repeatedly and \eqref{eq:coasso} in the third step.
This proves that $\rmat^{\bifund\fund}$ displays the expected cocommutativity properties.

\subsection{Auxiliary features}

Here we will discuss some auxiliary features of the fused R-matrices.

\paragraph{Similarity transformations.}

We have the freedom to apply a similarity transformation $W(u_\combine{12})$ to the space $\Vectors^\bifund$
\begin{align}\label{eq:similarity}
\embed(u_\combine{12})&\to\embed(u_\combine{12})\, W(u_\combine{12})^{-1},
\nln
\fuse(u_\combine{12})&\to W(u_\combine{12})\fuse(u_\combine{12}),
\nln
\rbody(u_\combine{12})&\to W(u_\combine{12})\rbody(u_\combine{12}) W(u_\combine{12})^{-1}.
\end{align}
This transformation affects none of the relations \eqref{eq:PpropertiesA},
and therefore all the above results apply to the transformed system
without further ado.
In the construction of $\embed,\rbody,\fuse$ via eigenvectors \eqref{eq:EHFeigen},
the similarity transformation
does not preserve the diagonal nature of $\rbody$
(unless $W$ is diagonal as well).
However, we have not made explicit use of this property in the constructions.

\paragraph{Opposite form.}

A similarity transformation by $\rbody_\combine{12}$ has a
curious effect on the fused R-matrices
\begin{align}
\label{eq:oppositeR}
\rbody_\combine{12}\rmat_{\combine{12}3}\rbody_\combine{12}^{-1}&=
\rbody_\combine{12}\fuse_{\combine{12}}
\rmat_{13}
\rmat_{23}
\embed_{\combine{12}}\rbody_\combine{12}^{-1}
\nln &=
\fuse_{\combine{12}} \rmat_{12}
\rmat_{13}
\rmat_{23}
\embed_{\combine{12}}\rbody_\combine{12}^{-1}
\nln &=
\fuse_{\combine{12}}
\rmat_{23}
\rmat_{13}
\rmat_{12}
\embed_{\combine{12}}\rbody_\combine{12}^{-1}
\nln &=
\fuse_{\combine{12}}
\rmat_{23}
\rmat_{13}
\embed_{\combine{12}}.
\end{align}
Compared to the original definition
$\rmat_{\combine{12}3}=
\fuse_{\combine{12}}
\rmat_{13}
\rmat_{23}
\embed_{\combine{12}}$,
we observe that conjugation by $\rbody_\combine{12}$
flips the order of the R-matrix factors
within the fused R-matrix.

This observation goes hand in hand with the definition \eqref{eq:DefFusedRep}
of the fused representation via the coproduct.
If instead of the coproduct we use the opposite coproduct,
the resulting representation
is related to the original one by a simple similarity transformation
\begin{align}
\fuse_\combine{12} \, \bigbrk{\rep_1 \otimes \rep_2} \bigsbrk{\copro(\gen{J})}\, \embed_\combine{12}
=
\rbody_\combine{12}^{-1}\, \fuse_\combine{12} \, \bigbrk{\rep_1 \otimes \rep_2} \bigsbrk{\coproop(\gen{J})}\,
\embed_\combine{12}\, \rbody_\combine{12}.
\end{align}
As usual we used \eqref{eq:Pproperties} and the intertwining property of the R-matrix.

\paragraph{Symmetric R-matrices.}

In many practical applications
$\rmat(u_\combine{12})$, acting as an operator on the space $\Vectors^\fund\otimes\Vectors^\fund$,
is symmetric w.r.t.\ some inner product,
e.g.\ the standard inner product $\langle a,b\rangle := a^{\trans} b$
defined on $\Vectors^\fund$.
In most cases,
$\rmat(u_\combine{12})=\rmat(u_\combine{12})^\trans$ admits an orthonormal basis of eigenvectors.%
\footnote{This is evident if the inner product is positive definite.
For indefinite inner products (including complex symmetric matrices),
eigenvectors can be null.
In this case the eigenvectors cannot be normalized,
and even a non-trivial Jordan decomposition may arise.}
In addition, the fusion and embedding matrices are conjugate to each other
$\fuse = \embed^\trans$.

One minor problem is that the resulting R-matrices
\eqref{eq:FusedRBF} are not symmetric.
Transposition reverses the order of the
constituent R-matrices which corresponds to
a similarity transformation according to \eqref{eq:oppositeR}
\begin{align}
\rmat_{\combine{12}3}^\trans
 =
\fuse_{\combine{12}} \rmat_{23} \rmat_{13} \embed_{\combine{12}}
=
\rbody_\combine{12} \rmat_{\combine{12}3} \rbody_{12}^{-1}.
\end{align}
By applying a similarity transformation \eqref{eq:similarity}
defined by a $W$ such that $\rbody = W^\trans W$,
the resulting R-matrices become symmetric
\[\label{eq:symmR}
\rmat_{\combine{12}3}'^\trans
=
(W\rmat_{\combine{12}3}W^{-1})^\trans
=
W^{-\trans}\rbody_\combine{12}\rmat_{\combine{12}3}\rbody_\combine{12}^{-1}W^\trans
=
(W^{-\trans}\rmat_{\combine{12}3}^\trans W^\trans)^\trans
=
\rmat_{\combine{12}3}'.
\]
%

\subsection{Complementary fusion}

A fused R-matrix can also be defined for the complement $\Vectors^\bifundcomp_{\combinecomp{12}}$
of the space $\Vectors^\bifund_{\combine{12}}$
\begin{align}
&\Vectors^\fund_{1}\otimes\Vectors^\fund_{2}
=
\Vectors^\bifund_{\combine{12}}
\oplus
\Vectors^\bifundcomp_{\combinecomp{12}},
&&
\rmat^{\bifundcomp\fund}_{\combinecomp{12}3}:
\Vectors^\bifundcomp_{\combinecomp{12}}
\otimes
\Vectors^\fund_{3}
\to
\Vectors^\bifundcomp_{\combinecomp{12}}
\otimes
\Vectors^\fund_{3}.
\end{align}
As we shall see, this space is even better suited for fusion.

\paragraph{Complementary space.}

To define an R-matrix for the complement, we need to define fusion and embedding matrices for the complementary space
\begin{align}
&\embedcomp(u_\combine{12}): \Vectors^{\bifundcomp} \rightarrow \Vectors^{\fund} \otimes \Vectors^{\fund},
&
&\fusecomp(u_\combine{12}): \Vectors^{\fund}\otimes\Vectors^{\fund} \rightarrow \Vectors^{\bifundcomp}.
\end{align}
They are defined to obey the following orthogonality properties
with the original embedding and fusion matrices in \eqref{eq:PpropertiesB}:
\begin{align}\label{eq:PpropertiesB}
&
\fusecomp(u_\combine{12})\, \embedcomp(u_\combine{12}) = 1^{\bifundcomp},
&&
\fuse(u_\combine{12})\, \embedcomp(u_\combine{12}) = 0,
&&
\fusecomp(u_\combine{12})\, \embed(u_\combine{12}) = 0.
\end{align}
They directly imply the completeness relations
\[\label{eq:completenessB}
\embed_\combine{12}\fuse_\combine{12}
+
\embedcomp_\combinecomp{12}\fusecomp_\combinecomp{12}=1_{12}.
\]
as well as orthogonality relations with the R-matrix
\begin{align}\label{eq:PpropertiesC}
&\rmat(u_\combine{12})\, \embedcomp(u_\combine{12}) =0 ,
&
&\fusecomp(u_\combine{12})\, \rmat(u_\combine{12})= 0 .
\end{align}

\paragraph{Complementary R-matrix.}

One can define a complementary R-matrix in analogy to \eqref{eq:FusedRBF}
\[\label{eq:FusedRBFcomp}
\rmat^{\bifundcomp\fund}_{\combinecomp{12}3}(u_{\combine{12}},u_3)  :=
\fusecomp_{\combinecomp{12}}(u_{\combine{12}}) \,
\rmat_{13}(u_1,u_3)\,
\rmat_{23}(u_2,u_3)\,
\embedcomp_{\combinecomp{12}}(u_{\combine{12}}).
\]
The other related R-matrices follow in a similar fashion.
The various integrability relationships can be derived
in a similar fashion to the above.
The general strategy is to remove the factors of
$\embedcomp_{\combinecomp{12}}\fusecomp_{\combinecomp{12}}$
which typically appear between the various R-matrices.
The starting point is the relationship
$\fuse_{\combine{12}}\rmat_{13}\rmat_{23}\embedcomp_{\combinecomp{12}}=0$
which follows from the above definitions
\[
\rbody_{\combine{12}}\fuse_{\combine{12}}
\rmat_{13}\rmat_{23}
\embedcomp_{\combinecomp{12}}
=
\fuse_{\combine{12}}
\rmat_{12}\rmat_{13}\rmat_{23}
\embedcomp_{\combinecomp{12}}
=
\fuse_{\combine{12}}
\rmat_{23}\rmat_{13}\rmat_{12}
\embedcomp_{\combinecomp{12}}
=0.
\]
Together with the completeness relationship \eqref{eq:completenessB}
one can show
\[
\embedcomp_{\combinecomp{12}}\fusecomp_{\combinecomp{12}}\rmat_{13}\rmat_{23}\embedcomp_{\combinecomp{12}}
=
\rmat_{13}\rmat_{23}\embedcomp_{\combinecomp{12}}
-\embed_{\combine{12}}\fuse_{\combine{12}}\rmat_{13}\rmat_{23}\embedcomp_{\combinecomp{12}}
=
\rmat_{13}\rmat_{23}\embedcomp_{\combinecomp{12}}.
\]
Therefore, all intermediate factors of
$\embedcomp_{\combinecomp{12}}\fusecomp_{\combinecomp{12}}$
can be removed iteratively
from products of R-matrices
from the left to the right.

Similarly, one can prove that
\begin{align}
&\fusecomp_{\combinecomp{12}}
\rmat_{23}\rmat_{13}
\embed_{\combine{12}}
=0,
&& \embed_{\combine{12}}\fuse_{\combine{12}}\rmat_{23}\rmat_{13}\embed_{\combine{12}}
= \rmat_{23}\rmat_{13}\embed_{\combine{12}}.
\end{align}
On the level of the R-matrix this simply corresponds to a similarity transformation
on the composite particle space applied to the original fused R-matrix \eqref{eq:oppositeR}.

Put differently, the combinations $\rmat_{13}\rmat_{23}$ and  $\rmat_{23}\rmat_{13}$,
when viewed as a block diagonal matrix,
effectively possess a triangular shape:
\begin{align}\label{eq:RBFtriangle}
&\rmat_{13}\rmat_{23}
:
\begin{cases}
\Vectors^\bifund_{\combine{12}}\otimes \Vectors^\fund_3
\to \bigbrk{\Vectors^\bifund_{\combine{12}} \oplus \Vectors^\bifundcomp_{\combinecomp{12}}}\otimes \Vectors^\fund_3,
\\
\Vectors^\bifundcomp_{\combinecomp{12}}\otimes \Vectors^\fund_3
\to \Vectors^\bifundcomp_{\combinecomp{12}}\otimes \Vectors^\fund_3.
\end{cases}
\nln
&\rmat_{23}\rmat_{13}
:
\begin{cases}
\Vectors^\bifund_{\combine{12}}\otimes \Vectors^\fund_3
\to \Vectors^\bifund_{\combine{12}} \otimes \Vectors^\fund_3,
\\
\Vectors^\bifundcomp_{\combinecomp{12}}\otimes \Vectors^\fund_3
\to \bigbrk{\Vectors^\bifund_{\combine{12}} \oplus \Vectors^\bifundcomp_{\combinecomp{12}}}\otimes \Vectors^\fund_3.
\end{cases}
\end{align}
Composite states from the complementary subspace $\Vectors^\bifundcomp_{\combinecomp{12}}$
are mapped to the same subspace by $\rmat_{13}\rmat_{23}$.
Conversely, composite states from the original subspace $\Vectors^\bifund_{\combine{12}}$
can map to both subspaces.
Fusion as defined in \eqref{eq:FusedRBF} works only
due to the presence of the projector $\fuse_{\combine{12}}$.
However, for the opposite R-matrix \eqref{eq:oppositeR}
this reverses and states from $\Vectors^\bifund_{\combine{12}}$
are mapped to the same subspace.
This is at the cost of a similarity transformation on $\Vectors^\bifund_{\combine{12}}$.

\paragraph{Opposite fusion.}

The involution property \eqref{eq:YBE} also has an interesting
implication on the complementary space $\Vectors^\bifundcomp$
which we shall discuss in the following.
According to our assumptions, the first $m$ eigenvalues of $\rmat(u_\combine{12})$
are non-zero while the other $n^2-m$ vanish at the point $u_\combine{12}=(u_1,u_2)$.
We assume that the latter fall off linearly in $\epsilon$ when approaching
the singular point as in $(u_1,u_2+\epsilon)\to u_\combine{12}$.
Furthermore, also the product $\rmat_{12}\rmat_{21}$
is assumed to be proportional to $\epsilon$.

By the involution property \eqref{eq:YBE}
we have that away from the singular points
$\rmat_{12}$ is invertible with inverse proportional to $\rmat_{21}$.
Therefore $\rmat_{12}$ and $\rmat_{21}$ share the same eigenvectors,
but with inverse eigenvalues.
Now, by our assumptions on the behavior near the singular point,
we are led to the conclusion that the first $m$ eigenvalues of $\rmat_{21}$
fall off linearly with $m$ while the remaining
$n^2-m$ eigenvalues approach a constant.%
\footnote{If some of the eigenvalues of $\rmat_{12}$ scale with a higher power of $\epsilon$
(consequently also $\rmat_{12}\rmat_{21}$),
only fewer than $n^2-m$ eigenvalues will be finite.
The following considerations would have to be adapted accordingly.
See \secref{sec:singlet} for a concrete example of this case.}

Therefore, the null space of $\rmat_{21}(u_\combine{21})$
at $u_\combine{21}:=(u_2,u_1)$
is given by the image of $\embed_\combine{12}(u_\combine{12})$
whereas the non-trivial eigenspace is the image of $\embed_\combine{21}(u_\combine{21})$.
Consequently, the space $\Vectors^{\bifundcomp}_\combine{21}$
is the complement of $\Vectors^{\bifund}_\combine{12}$ within
$\Vectors^{\fund}_1\otimes\Vectors^{\fund}_2$,
and it has dimension $\bar m:=n^2-m$.
We therefore reproduce the elements of complementary fusion
\begin{align}
&
\Vectors^\bifundcomp_{\combine{21}}=\Vectors^\bifundcomp_{\combinecomp{12}},
&&
\fuse_{\combine{21}}\sim\fusecomp_{\combinecomp{12}},
&&
\embed_{\combine{21}}\sim\embedcomp_{\combinecomp{12}}.
\end{align}
All the constructions for the complementary fused states
and operators proceed as before
with the roles of spaces $1$ and $2$ interchanged.

\subsection{Algebraic Bethe ansatz}

Let us briefly touch upon the effect of fusion on monodromy and transfer matrices
that play a central role in the algebraic Bethe Ansatz.

\paragraph{RTT-relation.}

The key relation in the algebraic Bethe ansatz is the so-called RTT-relation.
The RTT-relation describes the commutation relations
between the elements of an $n\times n$ dimensional operator valued matrix
$\tmat^{\fund}(u) : \Vectors^{\fund} \rightarrow \Vectors^{\fund} \otimes \mathcal{O}$,
called the monodromy matrix.
We have
\begin{align}
\rmat^{\fund\fund}_{12}(u_1,u_2) \tmat^{\fund}_1(u_1)  \tmat^{\fund}_2(u_2) =
\tmat^{\fund}_2(u_2)  \tmat^{\fund}_1(u_1) \rmat^{\fund\fund}_{12}(u_1,u_2).
\end{align}
However, rather than taking $\rmat^{\fund\fund}_{12}(u_1,u_2)$
one can also consider taking fused R-matrices
and consider the RTT relation this would generate.
To this end, we introduce a fused monodromy matrix
\begin{align}
\tmat^{\bifund}_\combine{12}(u_\combine{12}) :=
\fuse_\combine{12}(u_\combine{12}) \,
\tmat^{\fund}_1(u_1) \,
\tmat^{\fund}_2(u_2)\,
\embed_\combine{12}(u_\combine{12}).
\end{align}
It is straightforward to show that $\tmat^{\bifund}_\combine{12}$ satisfies the
RTT-relation for fused R-matrices.

An object of special interest is the transfer matrix,
which is defined as the trace of the monodromy matrix
\begin{align}
& t^{\fund} (u) = \tr_{\fund} \tmat^{\fund}(u),
&& t^{\bifund} (u_\combine{12})= \tr_{\bifund} \tmat^{\bifund}(u_\combine{12}).
\end{align}
This object generates the mutually commuting set of operators
that is the defining property of integrable systems.

\paragraph{Fusion for transfer matrices.}

We can formulate the relation
between the transfer matrices in different representations.
Consider the product of two transfer matrices
and use the completeness relation \eqref{eq:completenessB}
\begin{align}
t^{\fund}(u_1)\, t^{\fund}(u_2)
&= \tr_{12}\bigsbrk{\tmat_1(u_1) \tmat_2(u_2) } \nln
&=
\tr_\combine{12}\lrsbrk{\fuse_\combine{12}  \tmat_1(u_1) \tmat_2(u_2)\embed_\combine{12}}
+
\tr_\combine{21}\lrsbrk{\fuse_\combine{21} \tmat_1(u_1) \tmat_2(u_2)\embed_\combine{21}}
\nln
&=t^{\bifund}(u_\combine{12}) + t^{\bifundcomp}(u_\combine{21}).
\end{align}
For the latter term, we note that a similarity transformation by $\rbody_\combine{21}$ is required
to interchange the order of monodromy matrices in analogy to \eqref{eq:oppositeR}.

\section{The Heisenberg XXX spin chain}
\label{sec:XXX}

To illustrate our fusion procedure, let us first consider the Heisenberg XXX spin chain.
The fundamental particles transform in the spin-$\half$ representation of $\alg{su}(2)$.
The corresponding R-matrix is of difference form
$\rmat_{\text{XXX}}(u_1,u_2)=\rmat_{\text{XXX}}(u_1-u_2)$
and is given by
\begin{align}
\rmat_{\text{XXX}}(u) =
\begin{pmatrix}
u + 1 & 0 & 0 & 0 \\
0 & u & 1 & 0 \\
0 & 1 & u & 0 \\
0 & 0 & 0 & u + 1
\end{pmatrix}.
\end{align}
There are two points at which the rank of $\rmat_{\text{XXX}}(u)$ is not maximal,
namely at $u = \pm 1$.
At $u=1$ the rank is three, while at $u=-1$ the rank is one.
These points correspond to the singlet and triplet that arise
in the decomposition of the tensor product of two spin-$\half$ representations.

\paragraph{Singlet.}

At the point $u_1-u_2 = -1$,
there is only one eigenvector with non-zero eigenvalue and our projector is
\begin{align}
&\embed = \fuse^\trans =
\frac{1}{\sqrt{2}}
\begin{pmatrix}
0 \\
1 \\
-1 \\
0 \\
\end{pmatrix},
&&\rbody = -2.
\end{align}
It is easy to check that the identities from \eqref{eq:Pproperties} hold.
The representation that is generated according to \eqref{eq:DefFusedRep}
is the trivial representation.
Consequently, the fused R-matrix describing
the scattering of a singlet state with a fundamental particle is given by
\begin{align}
\rmat_{\combine{12}3}(u_\combine{12},u_3)
= \bigsbrk{(u_2-u_3)^2 -1} \begin{pmatrix}1&0\\0&1\end{pmatrix}.
\end{align}
In other words, we see that the singlet has trivial scattering
with a doublet
(up to an overall factor which depends on
the definition of the overall factor of $\rmat\indup{XXX}$).

\paragraph{Triplet.}

The other point $u_1-u_2=+1$ is the opposite of $u_1-u_2=-1$.
The resulting space therefore is the complement
of the above singlet.
More concretely, there is only one null vector
and from the three remaining eigenvectors we find
\begin{align}
&\embed = \fuse^\trans =
\begin{pmatrix}
1 & 0 & 0 \\
0 & \frac{1}{\sqrt{2}} &0 \\
0 & \frac{1}{\sqrt{2}} &0\\
0 & 0 & 1
\end{pmatrix},
&&\rbody = \diag(2,2,2).
\end{align}
It is easily seen that this will give rise to the usual spin-1 representation of $\alg{su}(2)$.
Indeed, \eqref{eq:DefFusedRep} yields for the simple roots
\begin{align}
&\rep_{\langle12\rangle}[\gen{S}^+] = \begin{pmatrix}
0 & \sqrt{2} & 0 \\
0 & 0 &\sqrt{2} \\
0 &0 &0
\end{pmatrix},
&&\rep_{\langle12\rangle}[\gen{S}^-] = \begin{pmatrix}
0 & 0 & 0 \\
\sqrt{2} & 0 & 0 \\
0 & \sqrt{2} &0
\end{pmatrix}.
\end{align}
Consequently, we recover the standard composite state R-matrix
$\rmat_{\combine{12}3}(u_\combine{12},u_3)$
for the triplet-doublet case
from \eqref{eq:FusedRBF}
\begin{align}
(u_1-u_3)
\begin{pmatrix}
 u_2-u_3+2 & 0 & 0 & 0 & 0 & 0 \\
 0 & u_2-u_3 & \sqrt{2} & 0 & 0 & 0 \\
 0 & \sqrt{2} &  u_2-u_3+1 & 0 & 0 & 0 \\
 0 & 0 & 0 & u_2-u_3+1  & \sqrt{2} & 0 \\
 0 & 0 & 0 & \sqrt{2} & u_2-u_3 & 0 \\
 0 & 0 & 0 & 0 & 0 &  u_2-u_3+2
\end{pmatrix}.
\end{align}
One can now easily check that the Yang--Baxter equation holds.

\section{The Hubbard model}
\label{sec:Hubb}

The R-matrix for the Hubbard model is best described in terms of its symmetry algebra;
centrally extended $\alg{su}(2|2)$.
This algebra is obtained from regular $\alg{su}(2|2)$
by adjoining two additional central charges to it.
The R-matrix in the fundamental representation is completely fixed
by the intertwining property \eqref{eq:symmS} \cite{Beisert:2005tm,Arutyunov:2006yd}.
The R-matrices involving composite state representations
are fixed by Yangian invariance \cite{deLeeuw:2008dp,Arutyunov:2009mi}.

It turns out that there are two cases where the R-matrix becomes of lower rank.
There are two cases where it becomes of rank 8
and the fused representation is a (a)symmetric short representation.
Moreover, there is a point where the R-matrix reduces to rank 1
corresponding to a singlet representation.

We will show that the bound state R-matrices
found in the literature  \cite{Arutyunov:2008zt} follow from our fusion procedure.

\subsection{The Hubbard model R-matrix}

The R-matrix for the Hubbard model is a $4^2\times 4^2$ matrix.
It acts on the tensor product of two $2|2$-dimensional spaces
with bosonic basis vectors $\state{\phi^{a}}$, $a=1,2$
and their fermionic counterparts
$\state{\psi^{\alpha}}$, $\alpha=3,4$.
For convenience let us introduce the $2|2$-dimensional
basis vector $E^A = (\state{\phi^{1}},\state{\phi^{2}},\state{\psi^{3}},\state{\psi^{4}})$
and let $E^A{}_B$ be the matrix unities with a $(-1)^{|B|}$ in row $A$ and column $B$.
The fundamental R-matrix is then of the form
\[\label{eqn:SinEE}
\rmat(u_1,u_2) = (-1)^{|B|+|C|} E^A{}_B\, \otimes E^C{}_D \rmat\ridx{B}{A}{D}{C}(u_1,u_2)
\]
with the only non-zero entries given by
\begin{align}
\label{eq:rmatbos}
 \rmat\ridx{a}{b}{c}{d}&=
\delta^a_d \delta^c_b +
(\delta^a_b \delta^c_d -\delta^a_d \delta^c_b )
\frac{x_1^+ - x_2^+}{x_1^- - x_2^+}\,
\frac{x_1^-}{x_1^+}\,
\frac{x_1^+x_2^- - 1}{x_1^-x_2^- -1}\,,
\nln
\rmat\ridx{\alpha}{\beta}{\gamma}{\delta} &=
\frac{\Ufund_2}{\Ufund_1}\,\frac{x^+_1-x^-_2}{x^-_1 - x^+_2}
\left[ \delta^\alpha_\delta \delta^\gamma_\beta +
(\delta^\alpha_\beta \delta^\gamma_\delta-\delta^\alpha_\delta \delta^\gamma_\beta)
\frac{x_1^+ - x_2^+}{x_1^+ - x_2^-}\,
\frac{x_2^-}{x_2^+}\,
\frac{x_1^-x_2^+ - 1}{x_1^-x_2^- -1}
\right],
\nln
\rmat\ridx{a}{\alpha}{b}{\beta} &=
-\varepsilon^{ab}\varepsilon_{\alpha\beta}
\frac{\gamma_1\gamma_2 \Ufund_2 x^-_1 x^-_2(x_1^+-x_2^+)}{x^+_1x^+_2(x^-_1x^-_2-1)(x^+_2 - x^-_1)}
\,,
\nln
 \rmat\ridx{\alpha}{a}{\beta}{b}& =
\varepsilon_{ab}\varepsilon^{\alpha\beta}
\frac{(x^+_1-x^+_2)(x^-_1-x^+_1)(x^-_2-x^+_2)}
{\gamma_1 \gamma_2 \Ufund_1(x^-_1 x^-_2-1)(x^+_2-x^-_1)}\,,
\end{align}
and
\begin{align}
\label{eq:rmatferm}
\rmat\ridx{a}{b}{\alpha}{\beta} &=\delta^a_b\delta^\alpha_\beta
\frac{1}{\Ufund_1}\,
\frac{x^+_1-x^+_2}{x^-_1-x^+_2}\,,
&\rmat\ridx{a}{\beta}{\alpha}{b} &= \delta^a_b\delta^\alpha_\beta
\frac{\Ufund_2}{\Ufund_1}\,
\frac{x^-_2-x^+_2}{x^+_2-x^-_1}\,\frac{\gamma_1}{\gamma_2}\,,
\nln
\rmat\ridx{\alpha}{b}{a}{\beta} &=\delta^a_b\delta^\alpha_\beta
\frac{x^+_1-x^-_1}{x^+_2-x^-_1}\,\frac{\gamma_2}{\gamma_1}\,,
&\rmat\ridx{\alpha}{\beta}{a}{b} &=\delta^a_b\delta^\alpha_\beta
\Ufund_2\,
\frac{x^-_1-x^-_2}{x^-_1-x^+_2}\,.
\end{align}
The parameters $x^\pm_{1,2}$ are related to the spectral parameters $u_{1,2}$,
respectively in the following way
\[\label{eqn:xpm2u}
u = x^+ + \frac{1}{x^+}  - \frac{\hbar}{2} = x^- + \frac{1}{x^-} + \frac{\hbar}{2}\,.
\]
The parameter $\hbar$ corresponds to the coupling constant
and the parameters $\Ufund$
are related to the above by $\Ufund^2 = x^+/x^-$.
Finally, the additional parameter $\gamma$
defines the relative normalization of bosons ($\phi$) and fermions ($\psi$).
%
The R-matrix is the intertwiner of the centrally extended $\alg{su}(2|2)$ superalgebra.
This algebra contains two $\alg{su}(2)$ subalgebras that act on the bosons and fermions respectively.
The R-matrix is a symmetric matrix with
respect to an appropriately chosen inner product for the states
\[\label{eq:inner22}
\langle \phi^a|\phi^b\rangle
=
\delta^{ab},
\qquad
\langle \psi^\alpha|\psi^\beta\rangle
=
\delta^{\alpha\beta}
\frac{\Ufund}{\gamma^2}\,\brk{x^{+}-x^{-}}
=
\delta^{\alpha\beta}
\frac{x^+}{\gamma^2}\,\brk{\Ufund-\Ufund^{-1}}.
\]

\subsection{Symmetric states}

Let us first consider the points where the R-matrix becomes of rank 8.
They correspond to the two special points $x_1^{+} = x_2^{-} $
(corresponding to $u_1 = u_2 + \hbar$) as well as $x_1^{-} = x_2^{+} $
(corresponding to $u_1 = u_2 - \hbar$).
In the following we will consider the point $x_1^+ = x_2^-$.
The considerations for the other point are completely analogous.

This first ingredient we need is the matrix of normalized eigenvectors $\embed$
of the fundamental R-matrix \eqref{eqn:SinEE}.
There are four bosonic vectors
$\state{\mathrm{S}^{(ab)}}$ and $\state{\mathrm{S}^{[34]}}$.
Three of them have unit eigenvalue $\lambda_{(ab)}=1$
and form a standard triplet under the bosonic $\alg{su}(2)$
\begin{align}
& \state{\mathrm{S}^{(aa)}} = \state{\phi^a} \otimes \state{\phi^a},
&&
\state{\mathrm{S}^{(12)}} = \frac{1}{\sqrt{2}}\,
\bigbrk{\state{\phi^1} \otimes \state{\phi^2} + \state{\phi^2} \otimes \state{\phi^1}},
\end{align}
and the fourth contains two fermionic states and is a singlet under both $\alg{su}(2)$'s
\begin{align}
\state{\mathrm{S}^{[34]}}
&=
\frac{\gamma_1\gamma_2 \Ufund_2^{-1}
 \bigsbrk{\varepsilon^{\alpha\beta}\state{\psi^\alpha} \otimes \state{\psi^\beta} }
+\bigbrk{1-\half(\Ufund_2^{-2} + \Ufund_1^2)}
 \bigsbrk{\varepsilon^{ab} \state{\phi^a} \otimes \state{\phi^b}}
}
{\sqrt{2}\sqrt{\Ufund_1 \Ufund_2^{-1}\brk{x^{+}_2-x^{-}_2}\brk{x^{+}_1-x^{-}_1}
+ \bigbrk{1-\half(\Ufund_2^{-2} + \Ufund_1^2)}^2}},
\nln
\lambda_{[34]} &=
\frac{x_2^+-x_2^-}{1-x_1^+x_1^-}
 \left[
 \frac{x^-_2 - x^-_1}{1-x^+_1x^+_2} +
\frac{2}{\Ufund_1\Ufund_2}\, \frac{1 - x_1^- x_2^+ }{x^+_2 - x^-_1}
\right].
\end{align}
There are four fermionic eigenstates $\state{\mathrm{S}^{a\alpha}}$
\begin{align}
&\state{\mathrm{S}^{a\alpha}}
=\frac{ \gamma_2\, \state{\phi^a} \otimes \state{\psi^\alpha}
+ \gamma_1 \Ufund_2\, \state{\psi^\alpha} \otimes \state{\phi^a} }
{\sqrt{\Ufund_2(x^+_2-x^-_2)+\Ufund_1\Ufund_2^2 (x^+_1-x^-_1)}}\,,
&& \lambda_{a\alpha} =\frac{1}{\Ufund_2}\,\frac{x_2^- - x_2^+}{x_1^- - x_2^+}
+\Ufund_1\frac{x_1^- - x_1^+}{x_1^- - x_2^+}.
\end{align}
After packaging these vectors in the matrix $\embed$ and taking $\fuse = \embed^\trans$,
it is readily checked that \eqref{eq:PpropertiesA} and \eqref{eq:Pproperties} hold.

We are now in a position to apply our fusion procedure.
Let us consider the fused matrix \eqref{eq:FusedRBF}
and make it symmetric by applying the transformation $W$ as in \eqref{eq:symmR}.
It is then straight-forward to check
that this R-matrix coincides with $\mathbb{S}^{BA}$ from \cite{Arutyunov:2008zt}.

As a non-trivial example let us compute two scattering processes
\begin{align}\label{eq:teststates}
&\rmat^{\prime BA}(u_\combine{12},u_3) \state{\mathrm{S}^{(11)}} \otimes \state{\phi^1},
&& \rmat^{\prime BA}(u_\combine{12},u_3) \state{\mathrm{S}^{13}} \otimes \state{\psi^3},
\end{align}
It is easy to see that
\begin{align}
&\rmat^{\prime BA}(u_\combine{12},u_3) \state{\mathrm{S}^{(11)}} \otimes \state{\phi^1}
= \state{\mathrm{S}^{(11)}} \otimes \state{\phi^1} ,
\end{align}
which agrees with $a_1=1$ from section 6.1.2 of \cite{Arutyunov:2008zt}.
This shows that we have the same normalization for the bound state S-matrix.
Subsequently, we have by definition
\begin{align}\label{eq:exampleRmat}
&\quad \rmat^{\prime BA} \state{\mathrm{S}^{13}} \otimes \state{\psi^3}
\nln
&=  W \fuse_{12} \rmat_{13} \rmat_{23} \embed_{12} W^{-1} \state{\mathrm{S}^{13}} \otimes \state{\psi^3}
\nln
&=
\biggl[
\frac{\Ufund_2 (x^+_1-x^-_1) (x^+_2-x^+_3)(x^-_3-x^+_1)}{(x^-_1-x^+_3)(x^+_3-x^-_2)}
+\frac{(x^+_2-x^-_2)(x^+_3-x^+_1)(x^-_3-x^+_2)}{\Ufund_1 (x^-_1-x^+_3)(x^-_2-x^+_3)}
 \nln & \qquad \quad
+\frac{\Ufund_2(x^+_2-x^-_2)(x^-_1-x^+_1)(x^-_3-x^+_3)}{(x^-_1-x^+_3)(x^+_3-x^-_2)}
\biggr]
\nln & \qquad
\cdot\frac{\Ufund_3}{\Ufund_2(x^+_2-x^-_2) + \Ufund_1\Ufund_2^2(x^+_1-x^-_1)} \,
\state{\mathrm{S}^{13}} \otimes \state{\psi^3}
\nln
&=\frac{\Ufund_3}{\Ufund_1\Ufund_2} \,
\frac{x^+_2-x^-_3}{x^-_1-x^+_3}\,
\state{\mathrm{S}^{13}} \otimes \state{\psi^3},
\end{align}
where we used the inner product \eqref{eq:inner22} and $x^+_1 = x^-_2$.
This result exactly coincides with the literature,
in particular it is the coefficient $a_7$ from section 6.1.2 of \cite{Arutyunov:2008zt}.
In fact it is not hard to check that the representation \eqref{eq:DefFusedRep}
is exactly the two particle bound state representation from \cite{Arutyunov:2008zt}.

Let us conclude this section by considering the complementary fusion procedure.
The complement is spanned by antisymmetric states $\state{\mathrm{A}}=
\set{\state{\mathrm{A}}^{(\alpha\beta)},\state{\mathrm{A}^{[12]}},\state{\mathrm{A}^{a\alpha}}}$
that are perpendicular
to the symmetric states $\state{\mathrm{S}}$ given above.

For these states, one can then easily compute the complementary R-matrix \eqref{eq:FusedRBFcomp}.
In particular, due to the upper triangular structure \eqref{eq:RBFtriangle},
the computation simplifies somewhat.
For instance, let us consider the analogue of the states \eqref{eq:teststates}
\begin{align}
& \state{\mathrm{A}^{(33)}} \otimes \state{\psi^3},
&&
\state{\mathrm{A}^{13}}  \otimes \state{\phi^1}.
\end{align}
The action of $\rmat_{13}\rmat_{23} $ on both states is simply multiplicative and yields
\begin{align}
\rmat_{13}\rmat_{23} \state{\mathrm{A}^{(33)}} \otimes \state{\psi^3}
&= \frac{x_1^+ - x^-_3}{x_1^- - x^+_3}\, \frac{U_3}{U_1} \cdot \frac{x_2^+ - x^-_3}{x_2^- - x^+_3}\,
 \frac{U_3}{U_2}\,  \state{\mathrm{A}^{(33)}} \otimes \state{\psi^3}  ,
\\
\rmat_{13}\rmat_{23} \state{\mathrm{A}^{13}} \otimes \state{\phi^1}
&=  \frac{x_1^+ - x^-_3}{x_1^- - x^+_3} \,\frac{U_3}{U_1} \cdot \frac{x_2^+ - x^-_3}{x_2^- - x^+_3} \,
\frac{U_3}{U_2} \cdot \frac{\Ufund_1 \Ufund_2}{\Ufund_3} \,
\frac{x^-_1-x^+_3}{x^+_2-x^-_3}\, \state{\mathrm{A}^{13}} \otimes \state{\phi^1}.
\end{align}
This then corresponds to the fused R-matrix on antisymmetric states corresponding
to the other point of lower rank $x^-_1 = x^+_2$.
Notice that the matrix has a different normalization
and the inverse of the coefficient \eqref{eq:exampleRmat} appears,
which is in agreement with \cite{Bajnok:2008bm}
where the relation between the symmetric and antisymmetric R-matrices is discussed.

\subsection{Singlet state}
\label{sec:singlet}

The R-matrix coefficients in \eqref{eq:rmatbos}
have a common factor of $(x_1^+-x_2^+)/(x^-_1x^-_2-1)$,
which has a potential singularity at $x_2^-=1/x_1^-$.%
\footnote{Here, we do not explicitly multiply the R-matrix by an overall
factor to compensate the singularity.
Hence we will consider the most singular contributions to the matrix.}
Furthermore, $x_2^+\neq x_1^+$ in order for the numerator to be non-zero,
i.e.\ $x_2^+=1/x_1^+$.
The singularity affects only the $\alg{su}(2)\times\alg{su}(2)$ singlet
sector spanned by the two states
\begin{align}
\ket{\mathrm{BB}} &=\varepsilon^{ab}\phi^a\otimes\phi^b,
&
\ket{\mathrm{FF}} &=\varepsilon^{\alpha\beta}\psi^\alpha\otimes\psi^\beta.
\end{align}
The action on the remaining 14 states is finite.
Acting on the singlet, the R-matrix reduces to a $2\times 2$ matrix $M$.
In the limit $x_2^\pm \to 1/x_1^\pm$,
it has the following singularity structure
\[
M =
\frac{1}{\epsilon}\,M^{(-1)}
+
M^{(0)}
+\ldots\,.
\]
Up to an overall factor we find for the residue
\[
M^{(-1)} \sim
\begin{pmatrix}
1
&
-(\Ufund_1-\Ufund_1^{-1})/\gamma_1\gamma_2
\\
\Ufund_1\Ufund_2\gamma_1\gamma_2/\brk{\Ufund_1-\Ufund_1^{-1}}
&
-\Ufund_1\Ufund_2
\end{pmatrix}.
\]
We know that $\Ufund^2=x^+/x^-$
and hence $\Ufund_1\Ufund_2=\pm 1$.
Both values of the latter signs are permitted,
and we have to discuss the two cases separately,
as they lead to rather distinct behavior.

We start with $\Ufund_1\Ufund_2=-1$
which is analogous to the cases discussed above.
There are two eigenvectors
\begin{align}
\label{eq:fermisinglets}
\brk{\Ufund_1-\Ufund_1^{-1}}
\ket{\mathrm{BB}}
+
\gamma_1\gamma_2
\ket{\mathrm{FF}}
&&\text{and}&&
\brk{\Ufund_1-\Ufund_1^{-1}}
\ket{\mathrm{BB}}
-
\gamma_1\gamma_2
\ket{\mathrm{FF}}.
\end{align}
The eigenvalues are $2$ and $0$, respectively,
therefore only the former state is singular.
It is a singlet of the Yangian algebra,
however it has a non-trivial charge
$\Ufund_{\combine{12}}=\Ufund_1\Ufund_2=-1$
which plays a role for the coproduct.
The other state belongs to an adjoint representation of $\alg{psu}(2|2)$.%
\footnote{The adjoint is reducible but indecomposable,
and the former singlet also acts as the top components of this representation.}
The singularity structure of this case is peculiar with respect to
complementary fusion: In the limit $x_2^\pm \to 1/x_1^\pm$,
the eigenvalue of the above non-singular singlet state
approaches zero even faster than for the 14 non-singlet states.
We find
\begin{align}
\lambda_{1\mathrm{a}}& \sim \frac{1}{\epsilon},
&
\lambda_{14} &\sim 1,
&
\lambda_{1\mathrm{b}} &\sim \epsilon.
\end{align}
This means that complementary fusion based on
$\rmat_{21}$ produces merely one composite state rather than 15.
This state is just the other singlet, and in fact one can see
that exchanging the two sites in \eqref{eq:fermisinglets}
interchanges the two states.%
\footnote{One has to take into account
the sign from exchanging fermions and
that $\Ufund_1\to-\Ufund_1^{-1}$.}

The other case $\Ufund_{\combine{12}}=\Ufund_1\Ufund_2=+1$ has
only one eigenvector
\[
\brk{\Ufund_1-\Ufund_1^{-1}}
\ket{\mathrm{BB}}
+
\gamma_1\gamma_2
\ket{\mathrm{FF}},
\]
and its eigenvalue is $0$.
The matrix $M^{(-1)}$ does not admit a second eigenvector because it has a
non-trivial Jordan decomposition. This makes the case very special,
and the fusion procedure described in this paper does not immediately apply.
Let us therefore try to understand what is going on:
The eigenstate is the singlet state discussed in \cite{Beisert:2005tm}.
There it was shown that the state behaves just like a fused state
under scattering with other states, i.e.\ it preserves its form.
To understand the role of the other singlet state, it makes sense to take a
closer look at the limit $x_2^\pm \to 1/x_1^\pm$.
Here, both singlet eigenvalues remain \emph{finite},
whereas the eigenvectors become \emph{collinear}.
Therefore, at $x_2^\pm =1/x_1^\pm$ there is only one meaningful eigenvector,
the difference of the eigenvectors plays no significant role.
Even though all eigenvalues remain finite at this point,
fusion does take place due to the coincidence of eigenvectors.

The fused R-matrix $\rmat_{\combine{12}3}$
of the singlet state with $\Ufund_{\combine{12}}=+1$
was shown to be trivial in \cite{Beisert:2005tm}
up to an overall phase factor related to crossing symmetry \cite{Janik:2006dc}.
Based on this result one can easily derive
the fused R-matrix of the singlet state with $\Ufund_{\combine{12}}=-1$.
The point is that the factor $\Ufund_1$ appears in odd powers
in \eqref{eq:rmatbos} and \eqref{eq:rmatferm} only
when the third index is fermionic.
Flipping the sign $\Ufund_2$
flips the sign of $\rmat_{\combine{12}3}$ precisely if
state $3$ is fermionic \eqref{eqn:SinEE}.
This means that the state with $\Ufund_{\combine{12}}=-1$
behaves like a fermionic singlet, while
$\Ufund_{\combine{12}}=+1$ corresponds to
a bosonic singlet. This observation is in line with the
coproduct rule of odd generators.

\pdfbookmark[1]{Acknowledgements}{ack}
\section*{Acknowledgements}

We would like to thank
G.\ Arutyunov, S.\ Frolov, J. Myers and A.\ Torrielli for discussions.
The work of NB and MdL is partially supported
by grant no.\ 200021-137616 from the Swiss National Science Foundation
and through the NCCR SwissMAP.
The work of NB is partially supported by grant no.\ 615203
from the European Research Council under the FP7. MdL was also supported by FNU through grant number
DFFÐ1323Ð00082.


\begin{bibtex}[\jobname]

@article{Beisert:2007ds,
      author         = "Beisert, Niklas",
      title          = "{The S-matrix of AdS / CFT and Yangian symmetry}",
      journal        = "PoS",
      volume         = "SOLVAY",
      pages          = "002",
      year           = "2006",
      eprint         = "0704.0400",
      archivePrefix  = "arXiv",
      primaryClass   = "nlin.SI",
      reportNumber   = "AEI-2007-019",
      SLACcitation   = "
}

@article{deLeeuw:2008dp,
      author         = "de Leeuw, Marius",
      title          = "{Bound States, Yangian Symmetry and Classical r-matrix
                        for the AdS$_5$ $\times$ S$^5$ Superstring}",
      journal        = "JHEP",
      volume         = "0806",
      pages          = "085",
      doi            = "10.1088/1126-6708/2008/06/085",
      year           = "2008",
      eprint         = "0804.1047",
      archivePrefix  = "arXiv",
      primaryClass   = "hep-th",
      reportNumber   = "ITP-UU-08-18, SPIN-08-17",
      SLACcitation   = "
}

@article{Arutyunov:2009mi,
      author         = "Arutyunov, Gleb and de Leeuw, Marius and Torrielli,
                        Alessandro",
      title          = "{The Bound State S-Matrix for AdS$_5$ $\times$ S$^5$ Superstring}",
      journal        = "Nucl.Phys.",
      volume         = "B819",
      pages          = "319-350",
      doi            = "10.1016/j.nuclphysb.2009.03.024",
      year           = "2009",
      eprint         = "0902.0183",
      archivePrefix  = "arXiv",
      primaryClass   = "hep-th",
      reportNumber   = "ITP-UU-09-06, SPIN-09-06",
      SLACcitation   = "
}

@article{Janik:2006dc,
      author         = "Janik, Romuald A.",
      title          = "{The AdS$_5$ $\times$ S$^5$ superstring worldsheet S-matrix and
                        crossing symmetry}",
      journal        = "Phys.Rev.",
      volume         = "D73",
      pages          = "086006",
      doi            = "10.1103/PhysRevD.73.086006",
      year           = "2006",
      eprint         = "hep-th/0603038",
      archivePrefix  = "arXiv",
      primaryClass   = "hep-th",
      SLACcitation   = "
}

@article{Kulish:1981gi,
      author         = "Kulish, P. P. and Reshetikhin, N. {\relax Yu}. and Sklyanin, E. K.",
      title          = "{Yang-Baxter Equation and Representation Theory. 1.}",
      journal        = "Lett.Math.Phys.",
      volume         = "5",
      pages          = "393-403",
      doi            = "10.1007/BF02285311",
      year           = "1981",
      SLACcitation   = "
}

@article{Martins:2007hb,
      author         = "Martins, M.J. and Melo, C.S.",
      title          = "{The Bethe ansatz approach for factorizable centrally
                        extended S-matrices}",
      journal        = "Nucl.Phys.",
      volume         = "B785",
      pages          = "246-262",
      doi            = "10.1016/j.nuclphysb.2007.05.021",
      year           = "2007",
      eprint         = "hep-th/0703086",
      archivePrefix  = "arXiv",
      primaryClass   = "hep-th",
      reportNumber   = "UFSCAR-TH-07-03",
      SLACcitation   = "
}

@article{Beisert:2006qh,
      author         = "Beisert, Niklas",
      title          = "{The Analytic Bethe Ansatz for a Chain with Centrally
                        Extended su(2$/$2) Symmetry}",
      journal        = "J.Stat.Mech.",
      volume         = "07",
      pages          = "P01017",
      doi            = "10.1088/1742-5468/2007/01/P01017",
      year           = "2007",
      eprint         = "nlin/0610017",
      archivePrefix  = "arXiv",
      primaryClass   = "nlin.SI",
      reportNumber   = "AEI-2006-074, PUTP-2211",
      SLACcitation   = "
}

@article{Beisert:2005tm,
      author         = "Beisert, Niklas",
      title          = "{The SU(2$/$2) dynamic S-matrix}",
      journal        = "Adv.Theor.Math.Phys.",
      volume         = "12",
      pages          = "945-979",
      year           = "2008",
      eprint         = "hep-th/0511082",
      archivePrefix  = "arXiv",
      primaryClass   = "hep-th",
      doi            = "10.4310/ATMP.2008.v12.n5.a1",
      reportNumber   = "PUTP-2181, NSF-KITP-05-92",
      SLACcitation   = "
}

@article{ShastryMatrix,
  year={1988},
  issn={0022-4715},
  journal={J. Stat. Phys.},
  volume={50},
  number={1-2},
  doi={10.1007/BF01022987},
  title={Decorated star-triangle relations and exact integrability of the one-dimensional Hubbard model},
  publisher={Kluwer Academic Publishers-Plenum Publishers},
  keywords={One-dimensional Hubbard model; exactly integrable systems; star-triangle relations},
  author={Shastry, B. Sriram},
  pages={57-79},
  language={English}
}

@article{Beisert:2010jr,
      author         = "Beisert, Niklas and others",
      title          = "{Review of AdS/CFT Integrability: An Overview}",
      journal        = "Lett.Math.Phys.",
      volume         = "99",
      pages          = "3-32",
      doi            = "10.1007/s11005-011-0529-2",
      year           = "2012",
      eprint         = "1012.3982",
      archivePrefix  = "arXiv",
      primaryClass   = "hep-th",
      reportNumber   = "AEI-2010-175, CERN-PH-TH-2010-306, HU-EP-10-87,
                        HU-MATH-2010-22, KCL-MTH-10-10, UMTG-270, UUITP-41-10",
      SLACcitation   = "
}

@article{Arutyunov:2006ak,
      author         = "Arutyunov, Gleb and Frolov, Sergey and Plefka, Jan and
                        Zamaklar, Marija",
      title          = "The Off-Shell Symmetry Algebra of the Light-Cone AdS$_5$ $\times$ S$^5$",
      journal        = "J.Phys.",
      volume         = "A40",
      pages          = "3583-3606",
      doi            = "10.1088/1751-8113/40/13/018",
      year           = "2007",
      eprint         = "hep-th/0609157",
      archivePrefix  = "arXiv",
      primaryClass   = "hep-th",
      reportNumber   = "AEI-2006-071, HU-EP-06-31, ITP-UU-06-39, SPIN-06-33,
                        TCDMATH-06-13",
      SLACcitation   = "
}

@article{Arutyunov:2006yd,
      author         = "Arutyunov, Gleb and Frolov, Sergey and Zamaklar, Marija",
      title          = "{The Zamolodchikov-Faddeev algebra for AdS$_5$ $\times$ S$^5$
      superstring}",
      journal        = "JHEP",
      volume         = "0704",
      pages          = "002",
      doi            = "10.1088/1126-6708/2007/04/002",
      year           = "2007",
      eprint         = "hep-th/0612229",
      archivePrefix  = "arXiv",
      primaryClass   = "hep-th",
      reportNumber   = "AEI-2006-099, ITP-UU-06-58, SPIN-06-48, RCDMATH-06-18",
      SLACcitation   = "
}

@article{Dorey:2006dq,
      author         = "Dorey, Nick",
      title          = "{Magnon Bound States and the AdS/CFT Correspondence}",
      journal        = "J.Phys.",
      volume         = "A39",
      pages          = "13119-13128",
      doi            = "10.1088/0305-4470/39/41/S18",
      year           = "2006",
      eprint         = "hep-th/0604175",
      archivePrefix  = "arXiv",
      primaryClass   = "hep-th",
      SLACcitation   = "
}

@article{Chen:2006gea,
      author         = "Chen, Heng-Yu and Dorey, Nick and Okamura, Keisuke",
      title          = "{Dyonic giant magnons}",
      journal        = "JHEP",
      volume         = "0609",
      pages          = "024",
      doi            = "10.1088/1126-6708/2006/09/024",
      year           = "2006",
      eprint         = "hep-th/0605155",
      archivePrefix  = "arXiv",
      primaryClass   = "hep-th",
      reportNumber   = "DAMTP-06-38",
      SLACcitation   = "
}

@article{Arutyunov:2008zt,
      author         = "Arutyunov, Gleb and Frolov, Sergey",
      title          = "{The S-matrix of String Bound States}",
      journal        = "Nucl.Phys.",
      volume         = "B804",
      pages          = "90-143",
      doi            = "10.1016/j.nuclphysb.2008.06.005",
      year           = "2008",
      eprint         = "0803.4323",
      archivePrefix  = "arXiv",
      primaryClass   = "hep-th",
      reportNumber   = "ITP-UU-08-15, SPIN-08-14, TCDMATH-08-03",
      SLACcitation   = "
}

@article{Kirillov:1987zz,
      author         = "Kirillov, A. N. and Reshetikhin, N. {\relax Yu}.",
      title          = "{Exact solution of the integrable XXZ Heisenberg model
                        with arbitrary spin. I. The ground state and the
                        excitation spectrum}",
      journal        = "J.Phys.",
      volume         = "A20",
      pages          = "1565-1585",
      doi            = "10.1088/0305-4470/20/6/038",
      year           = "1987",
      SLACcitation   = "
}
@article{Jimbo:1985vd,
      author         = "Jimbo, Michio",
      title          = "{A q Analog of U(gl(n+1)), Hecke Algebra and the
                        Yang-Baxter Equation}",
      journal        = "Lett.Math.Phys.",
      volume         = "11",
      pages          = "247",
      doi            = "10.1007/BF00400222",
      year           = "1986",
      reportNumber   = "RIMS-517",
      SLACcitation   = "
}

@article{Jimbo:1985zk,
      author         = "Jimbo, Michio",
      title          = "{A q difference analog of U(g) and the Yang-Baxter
                        equation}",
      journal        = "Lett.Math.Phys.",
      volume         = "10",
      pages          = "63-69",
      doi            = "10.1007/BF00704588",
      year           = "1985",
      SLACcitation   = "
}

@article{Mezincescu:1991ke,
      author         = "Mezincescu, Luca and Nepomechie, Rafael I.",
      title          = "{Fusion procedure for open chains}",
      journal        = "J.Phys.",
      volume         = "A25",
      pages          = "2533-2544",
      year           = "1992",
      doi            = "10.1088/0305-4470/25/9/024",
      reportNumber   = "CERN-TH-6152-91, UMTG-163",
      SLACcitation   = "
}

@article{Karowski:1978ps,
      author         = "Karowski, M.",
      title          = "{On the Bound State Problem in (1+1)-dimensional Field
                        Theories}",
      journal        = "Nucl.Phys.",
      volume         = "B153",
      pages          = "244",
      doi            = "10.1016/0550-3213(79)90600-X",
      year           = "1979",
      reportNumber   = "FUB-HEP 27/78",
      SLACcitation   = "
}

@article{Arutyunov:2007tc,
      author         = "Arutyunov, Gleb and Frolov, Sergey",
      title          = "{On String S-matrix, Bound States and TBA}",
      journal        = "JHEP",
      volume         = "0712",
      pages          = "024",
      doi            = "10.1088/1126-6708/2007/12/024",
      year           = "2007",
      eprint         = "0710.1568",
      archivePrefix  = "arXiv",
      primaryClass   = "hep-th",
      reportNumber   = "ITP-UU-07-50, SPIN-07-37, TCDMATH-07-15",
      SLACcitation   = "
}

@book{BaxterBook,
    Author = {Rodney J. Baxter},
    Title = {Exactly solved models in statistical mechanics},
    Year = {1982},
    Language = {English},
    publisher = {Academic Press},
    address = {London, UK},
    HowPublished = {New York etc.: 486 p.},
}

@article{Bajnok:2008bm,
      author         = "Bajnok, Zoltan and Janik, Romuald A.",
      title          = "{Four-loop perturbative Konishi from strings and finite
                        size effects for multiparticle states}",
      journal        = "Nucl.Phys.",
      volume         = "B807",
      pages          = "625-650",
      doi            = "10.1016/j.nuclphysb.2008.08.020",
      year           = "2009",
      eprint         = "0807.0399",
      archivePrefix  = "arXiv",
      primaryClass   = "hep-th",
      SLACcitation   = "
}
\end{bibtex}

\bibliographystyle{nb}
\bibliography{\jobname}

\end{document}